\documentclass{ws-ijmpb}
\usepackage{color} 
\usepackage{url}
\usepackage{epsfig}
\usepackage{graphicx}
\usepackage{exscale}
\usepackage{lscape}
\def\ft{\hspace{0.1cm}}

\def\an1{$n^{(1)}(r)$}
\def\ea{{\it et al.}}
\def\bn2{$n^{(2)}(r)$}
\def\rb85{${}^{85}$Rb}
\def\cn01{$n_0^{(1)}(r)$}
\def\Rb87{${}^{87}$Rb}
\def\hs{\hspace{0.4cm}}
\def\dn02{$n_0^{(2)}(r)$}
\def\bls{\baselineskip -0.2cm}
\def\nac3{$na_c^3$}
\def\xna3{$na_{HS}^3$}
\def\n0{$n_0$}
\def\cn01{$n_0^{(1)}$}
\def\dn02{$n_0^{(2)}$}
\def\ahoeq{$a_{ho}\,=\,\sqrt{\hbar/m\omega_{ho}}$}
\def\li7{Li$^7$}
\def\bib{B\kern-.05em{I}\kern-.025em{B}\kern-.08em}
\def\btex{B\kern-.05em{I}\kern-.025em{B}\kern-.08em\TeX}
\begin{document}

\title{Thermodynamic properties of an interacting hard-sphere Bose gas in a trap
using the static fluctuation approximation}
\author{Saleem I. Qashou}
\address{Department of Physics, Faculty of Science, University of Jordan, 
Amman, JORDAN}
\author{Mohamed K. Al-Sugheir}
\address{Department of Physics, The Hashemite University, Zarqa, JORDAN}
\author{Asaad R. Sakhel}
\address{Faculty of Engineering Technology, Al-Balqa Applied University 
Amman 11134, JORDAN}
\author{Humam B. Ghassib}
\address{Department of Physics, University of Jordan, Amman, JORDAN}
\date{\today}
\maketitle

\begin{abstract}
\footnotesize\bls A hard-sphere (HS) Bose gas in a trap is investigated at 
finite temperatures in the weakly-interacting regime and its thermodynamic 
properties are evaluated using the static fluctuation approximation (SFA). The 
energies are calculated with a second-quantized many-body Hamiltonian and a 
harmonic oscillator wave function. The specific heat capacity, internal energy, 
pressure, entropy and the Bose-Einstein (BE) occupation number of the system are 
determined as functions of temperature and for various values of interaction strength 
and number of particles. It is found that the number of particles plays a more 
profound role in the determination of the thermodynamic properties of the system 
than the HS diameter characterizing the interaction, that the critical temperature 
drops with the increase of the repulsion between the bosons, and that the fluctuations 
in the energy are much smaller than the energy itself in the weakly-interacting regime.
\end{abstract}


\section{Introduction}

\hs The thermodynamic properties of a trapped interacting Bose gas have 
been investigated extensively in the last decade 
\cite{Holzmann:2004,Kim:1999,Li:2007,Grossmann:1995,Ohberg:1997,Grether:2003},
ever since Bose-Einstein condensation (BEC) was realized for the first time 
in a trap in 1995 \cite{Anderson:1995}. Finite-size effects \cite{Liu:2000} 
and the possibility of BEC in one and two dimensions \cite{Ketterle:1996} have 
been explored. Numerical \cite{Ohberg:1997} studies of interacting BECs in traps as
well as quasi one- and two-dimensional condensate expansion \cite{Kamchatnov:2004} 
due to the interactions have been carried out. In particular, the effect of the 
interaction on the transition temperature $T_c$ \cite{Giorgini:1996,Kao:2006,Holzmann:2004}
has been probed extensively. Further, the noninteracting trapped Bose gas has been a 
fascinating research topic 
\cite{Kao:2003,Rochin:2005,Napolitano:1997,Grossmann:1995,Gnanapragasam:2006} and its 
thermodynamic properties such as the BE occupation function, specific 
heat capacity, internal energy, and entropy have been calculated.
A basic issue in this context is the role of the energy fluctuations. For example, 
Giorgini \cite{Giorgini:2000} found that, below $T_c$, these fluctuations 
cause a shift in the excitation frequencies of order of a few percent.

\hs The aim of this paper is to undertake a full investigation of the 
thermodynamic properties for a trapped hard-sphere (HS) Bose gas with small 
and large numbers of particles by using a discrete-states approach. The role 
of the energy fluctuations in determining these properties will be underlined.

\hs We use the static fluctuation approximation (SFA) 
\cite{Joudeh:2005,Sandouqa:2006,Nigmatullin:2000} which includes in a relatively 
simple manner the above energy fluctuations. The principal feature of SFA is the 
replacement of the square of the local energy fluctuation operator 
$(\Delta \hat{E})^2$ with its mean value $\langle (\Delta \hat{E})^2 \rangle$. 
$[\langle (\Delta \hat{E})^2 \rangle]^{1/2}$ turns out to be much less than the 
chemical potential $\mu$ which, according to Giorgini \cite{Giorgini:2000}, 
places the fluctuations in our system in the collective-mode category. SFA has 
been used to study liquid $^{4}$He \cite{Al-Sugheir:2001}, liquid $^{3}$He 
\cite{Al-Sugheir:2002}, spin-polarized atomic hydrogen \cite{Joudeh:2005}, 
spin-polarized ${}^3$He-He II mixtures \cite{Sandouqa:2006}, ferroelectrics 
\cite{Nigmatullin:2000}, and nuclear matter \cite{Ghulam:2007}. All these are 
homogeneous systems. In contrast, SFA is used here for the first time to 
investigate an inhomogeneous-density system: a trapped Bose gas whose 
inhomogeneiety is a result of external trapping. The non-uniform Bose condensed 
gas has been investigated by \"Ohberg and Stenholm \cite{Ohberg:1997} in the 
framework of a Hartree-Fock (HF) formulation. However, HF theory fails at 
temperatures near the critical temperature $T_c$, whereas SFA works well below 
$T_c$. Hence our present SFA formulation of the trapped Bose gas may be regarded 
as a complementary method to that of \"Ohberg and Stenholm. Further, an approach is 
invoked according to which the particles are distributed in discrete energy 
levels; whereas most previous work, e.g. 
\cite{Liu:2000,Grossmann:1995,Sevincli:2007,Yukalov:2006}, used a continuous density of states. 

\hs We find that thermal fluctuations, for the ground state of a BEC in a 
harmonic trap, reach a maximum at the BEC transition temperature, 
and decline afterwards as the temperature $T$ rises. In addition, we find that at $g\,<\,1\times 10^{-3}$ the energy 
fluctuations are very small $\sim$O($10^{-5}$) and do not influence the thermodynamic properties very much. 
The entropy is found to increase with $T$ up to $T_c$, beyond which it begins to stabilize, reaching 
a plateau at higher $T$. Hence, the highest order in the system occurs when $T\rightarrow 0$; whereas 
at higher $T$ disorder is independent of $T$. It turns out that the critical temperature for the onset 
of BEC decreases with the increase of the repulsion between the bosons. It is further
found that a change in the number of particles of the system has quite a strong
influence on the thermodynamic properties. The Bose-Einstein (BE) occupation function is
studied as a function of $T$ and the effect of interaction on the distribution of the 
bosons in different HO states is investigated. We find that indeed a large number of discrete HO states
is needed ($\gg 100$) to describe strongly-repulsive BECs.

\hs The paper is organized as follows. In Sec.\ref{sec:method} the method is presented
briefly. In Sec. \ref{sec:analysis-of-sfa} the performance of an updated version of the SFA 
code is analyzed. In Sec. \ref{sec:results-and-discussion} our results are discussed and 
compared to previous literature. In Sec. \ref{sec:conclusions} our conclusions are listed. 
Appendix \ref{sec:sfa-iterative-procedure} explains the SFA iterative procedure used, and 
Appendix \ref{sec:sfa} describes the SFA method.

\section{Method}\label{sec:method}

\hs We consider an interacting hard-sphere (HS) Bose gas of $N$ particles
in an isotropic, spherically symmetric, harmonic trap of trapping frequency $\omega_{ho}\,=\,2\pi \times 10$ Hz and at finite temperature $T$. The strength of the interatomic interaction is 
characterized by the s-wave scattering length $a_s$, which is equivalent to 
the HS boson diameter in the low-energy and long-wavelength approximation.
We consider a hard core repulsive interaction between the bosons. 
In what follows, we present only the key points of the method; details are 
relegated to Appendices \ref{sec:sfa-iterative-procedure} and \ref{sec:sfa}. 

\subsection{SFA}

\hs Here we shed light on the small parameter of our approximation. As stated in the Introduction, we 
replace the square of the local energy fluctuation operator $(\Delta \hat{E}_m)^2$ for each state $m$ with its
mean value $\langle\Delta \hat{E}_m\rangle^2$. The energy fluctuation $\Delta \hat{E}_m$ is given by

\begin{equation}
\Delta \hat{E}_m\,=\,\hat{E}_m\,-\,\langle \hat{E}_m \rangle,
\end{equation}

where $\hat{E}_m$ is the energy operator of state $m$ [Eq.(\ref{eq:Em_n1n3})], and $\langle \hat{E}_m \rangle$ is the average value [Eq.(\ref{eq:avenergy_m})]. We define the small parameter of the approximation,
written $\varphi_F(m,T)$, which appears in the thermodynamic functions in Sec.\ref{sec:thermodynamic-functions} below, as

\begin{equation}
\varphi_F^2(m,T)\,=\,\langle (\Delta \hat{E}_m)^2\rangle.
\end{equation}

That is, 

\begin{equation}
\langle (\Delta \hat{E}_m)^2\rangle\,=\,\langle \hat{E}_m^2 \rangle\,-\,
\langle \hat{E}_m \rangle^2,
\end{equation}

as usual. Essentially, $\varphi_F(m,T)$ is proportional to the magnitude of 
the interaction constant $g$ as well as the average correlations between pairs 
of number fluctuations [see Eq.(\ref{eq:phi(m,T)^2}) below].

\subsection{Hamiltonian}\label{sec:hamiltonian}

\hs To evaluate the energy of the system, we use the following mean-field (MF) 
approach. We begin with the general form of the Hamiltonian:

\begin{eqnarray}
&&\hat{H}\,=\,\nonumber\\
&&\displaystyle\int d\mathbf{r} \hat{\psi}^\dagger(\mathbf{r})\left[-\frac{\hbar^2}{2 m}\nabla^2\,+\,V_{ext}(\mathbf{r})\right]\hat{\psi}(\mathbf{r})\,+\,
\frac{1}{2}\,\displaystyle\int d\mathbf{r}\,\hat{\psi}^\dagger(\mathbf{r})\hat{\psi}^\dagger(\mathbf{r})V_{int}
(\mathbf{r}-\mathbf{r}^\prime)\hat{\psi}(\mathbf{r})\hat{\psi}(\mathbf{r}). \nonumber\\
\label{eq:general-Hamiltonian}
\end{eqnarray}

To describe the interaction between the bosons in the trap, we first consider 
a two-body HS contact potential given by

\begin{equation}
V_{int}(\mathbf{r}_1-\mathbf{r}_2)\,=\,g\delta(\mathbf{r}_1-\mathbf{r}_2), 
\label{eq:Vint}
\end{equation}
where $g\equiv 4\pi\hbar^2a_s/u$ is the interaction parameter, with $a_s$ the s-wave
scattering length in the low-energy and long-wavelength approximation and $u$ 
the reduced bosonic mass. We further write the field operators as linear combinations of 
creation and annihilation operators:

\begin{equation}
\hat{\psi}(\mathbf{r})\,=\,\sum_m \phi_m(\mathbf{r})\hat{b}_m, 
\label{eq:field-operator}
\end{equation}

$\phi_n(\mathbf{r})$ being single-particle noninteracting wavefunctions for 
a particle at position $\mathbf{r}$ and $\hat{b}_m$ is the boson annihilation operator.

\hs Substituting (\ref{eq:Vint}) and (\ref{eq:field-operator}) into 
(\ref{eq:general-Hamiltonian}) and taking for $\phi_n(\mathbf{r})$ the harmonic 
oscillator wavefunctions in Cartesian coordinates:

\begin{equation}
\phi_m(\mathbf{r})\,=\,\phi_{m_x}(x)\phi_{m_y}(y)\phi_{m_z}(z), 
\label{eq:phixyz}
\end{equation}

with

\begin{equation}
\phi_m(x)\,=\,\frac{1}{2^m m! \sqrt{\pi}}\,\exp(-x^2/2) H_m(x), 
\label{eq:HOwavefunction}
\end{equation}

where $H_m(x)$ is the Hermite function of order $m$ (and similarly for $y$ and $z$), 
one gets for the second-quantized Hamiltonian for $N$ interacting bosons in a harmonic trap 

\begin{eqnarray}
&&\hat{H}\,=\,\displaystyle\sum_m \hbar\omega_{ho}(m+3/2)\,\hat{b}_m^\dagger\hat{b}_m\,+ \nonumber\\ 
&&\frac{g}{2}\,\sum_{c_1 c_2 c_3 c_4} \hat{b}_{c_1}^\dagger
\hat{b}_{c_2}^\dagger \hat{b}_{c_3} \hat{b}_{c_4} 
\int \phi_{c_1}^*(\mathbf{r}) \phi_{c_2}^*(\mathbf{r}) \phi_{c_3}(\mathbf{r}) 
\phi_{c_4}(\mathbf{r}) d\mathbf{r},  \label{eq:HamiltonianSQ}
\end{eqnarray}

$m$ and $c_i$ being integers representing harmonic oscillator (HO) states. 
Then, as shown in Appendix \ref{sec:sfa} [from Eq.(\ref{eq:commut.H0.bdagger}) 
to (\ref{eq:average-energy-incomplete})], the MF average energy per state $m$ 
can be written:

\begin{equation}
\langle \hat{E}_m \rangle\,=\,\hbar\omega_{ho}(m+3/2)\,+\,\frac{1}{2}g\sum_{n=0}^\infty c(n,m) \langle \hat{n}_k \rangle, \label{eq:avenergy_m}
\end{equation}

where $\langle \hat{n}_k \rangle$ is the BE occupation function:

\begin{equation}
\langle \hat{n}_k \rangle\,=\,\frac{d_k}{\exp[(E_k-\mu)\beta]-1}, 
\label{eq:Bose-Einstein.occup}
\end{equation}

with $\beta\,\equiv\,1/k_B T$, $k_B$ being Boltzmann's constant, $d_k$ a statistical weight defined
by the degeneracy of the system, and $c(k,m)$ the elements of the interaction matrix obtained from
the integral in Eq.(\ref{eq:HamiltonianSQ}):

\begin{eqnarray}
&&c(n,m)\,=\,\frac{1}{d_m}\sum_{\stackrel{\displaystyle n_x,n_y,n_z}
{\displaystyle m_x,m_y,m_z}} \prod_{i=1}^3 
\displaystyle \left[B_{n_i,m_i}
\int_{-\infty}^{+\infty} \exp(-2u_i^2) H_{n_i}^2(u_i) 
H_{m_i}^2(u_i) d u_i\right].
\label{eq:interaction-matrix-Cnm}
\end{eqnarray}

Here $m$ must fullfil the condition $m=m_x+m_y+m_z$, which leads to a degeneracy in each state,
and $d_m\,\equiv\,(m+1)(m+2)/2$ is the degeneracy of state $m$: $m_1=m_x$, $m_2=m_y$, $m_3=m_z$, 
and similarly for $n$; $u_1=x$, $u_2=y$, $u_3=z$. The reason $c(n,m)$ has been divided by the
degeneracy $d_m$ is that Eq.\ft(\ref{eq:interaction-matrix-Cnm}) is a sum over all possible
sets of degenerate HO states $(m_x,m_y,m_z)$ satisfying $m=m_x+m_y+m_z$ and interacting with all
possible sets of degenerate states $(n_x,n_y,n_z)$ with $n=n_x+n_y+n_z$. However, Eq.\ft(\ref{eq:avenergy_m}) describes a particle in a given state $m$ for a single set 
$(m_x,m_y,m_z)$ interacting with all the other sets $(n_x,n_y,n_z)$. Therefore, to ensure 
correct counting, $c(n,m)$ must be divided by $d_m$. $B_{n_i,m_i}$ stands for the normalization 
factor for each direction $i$, and is given by

\begin{equation}
B_{n_i,m_i}\,=\,\frac{1}{2^{n_i+m_i}\,n_i! m_i! \pi}.\label{eq:normalization-factor-x}
\end{equation}

Our computational limitations have restricted the rank of the matrix 
$c(n, m)$, i.e., the maximum values of $n$ and $m$.

\hs We use trap units for the energy and length, $\hbar\omega_{ho}$ and \ahoeq, 
respectively. Therefore, 

\begin{equation}
g\,\rightarrow\,\frac{g}{\hbar\omega_{ho}}\,=\,4\pi a_s a_{ho}^2; 
\label{eq:g_in_trap_units}
\end{equation}

and similarly $E\,\rightarrow\,E/\hbar\omega_{ho}$; by multiplying $g$ by a factor $a_{ho}^{-3}$ arising from the triple 
integral of the interaction matrix, $g$ becomes $4\pi a_s$ in trap units ($a_s/a_{ho}\rightarrow a_s$).
\subsection{Numerics}

\hs The integrals in Eq.\,(\ref{eq:interaction-matrix-Cnm}) are determined by 
applying a standard Gaussian quadrature technique \cite{Press:1999}. The 
Hermite functions are evaluated by using the recursion formula \cite{Arfken:1995} 

\begin{equation}
H_{n+1}(x)\,=\,2xH_n(x)\,-\,2nH_{n-1}(x); \label{eq:Hermite-recursion-formula}
\end{equation}

with $H_0(x)=1$ and $H_1(x)=2x$, one can iterate the 
above formula to get $H_n(x)$ for any $n$.
\subsection{Chemical potential}\label{sec:chemical-potential}

\hs In what follows a new, simple technique is presented that facilitates 
a speedy and accurate evaluation of the chemical potential $\mu$ for each temperature 
$T$. Since the sum of $\langle \hat{n}_k \rangle$ [Eq.(\ref{eq:Bose-Einstein.occup})] 
for a given $\mu$ yields the total number of particles, we define the function

\begin{equation}
f(\mu,T)\,=\,\left\{N\,-\,\sum_{m=0}^M \frac{d_m}{\exp[(\langle E_m 
\rangle\,-\,\mu)\beta]\,-\,1}\right\}^2, \label{eq:minimization-function}
\end{equation}

which has been minimized at each $T$ so as to tune $\mu(T)$ to a desired number of 
particles $N$ by using Powell's minimization routine \cite{Press:1999}. Here $M$ is 
the total number of states used. However, for the evaluation of $\mu$ 
in SFA, the `corrected' occupation function $\eta_0(m,T)$ weighted by $d_m$, which 
contains the fluctuations $\varphi_F$ [Eq.\,(\ref{eq:eta0})], has been used instead 
of $\langle \hat{n}_m \rangle$ [Eq.\,(\ref{eq:Bose-Einstein.occup})]. The evaluation 
of $\varphi_F$ using the SFA iterative procedure is explained in Appendix 
\ref{sec:sfa-iterative-procedure}.

\hs Thus, one first evaluates $\mu$ for the noninteracting system by setting 
$\langle E_m \rangle\,=\,\hbar\omega_{ho}(m+3/2)$ in Eq.\ft(\ref{eq:minimization-function}); 
next this $\mu$ is used in the evaluation of $\langle E_m \rangle$ for an 
interacting system via Eq.\,(\ref{eq:avenergy_m}). The energy of 
this interacting system is then fed back into Eq.\,(\ref{eq:minimization-function}) 
to get $\mu$ for the interacting system. This technique is 
somewhat similar to that used by Napolitano \ea\ \cite{Napolitano:1997}.
$M$ is tuned further to get the correct value for the specific heat 
capacity in the classical limit (Sec. \ref{sec:Cv-properties}).

\subsection{Thermodynamic functions}\label{sec:thermodynamic-functions}

\hs In this section the equations used in the SFA program are reviewed briefly.
The pressure is evaluated in trap units using 

\begin{equation}
P\Omega\,=\,\frac{k_B T}{\hbar\omega_{ho}} \ln Q, \label{eq:sfa-pressure}
\end{equation}

where $\Omega$ is the volume of the system, $Q$ is the Gibbs partition 
function, and

\begin{equation}
\ln Q \,=\, -\sum_{m=0}^M q_0(m,T), \label{eq:sfa-partition-function}
\end{equation}

$q_0(m,T)$ being a discrete function given by Eq.(\ref{eq:q_0(m,T)}).

\hs According to Rochin \cite{Rochin:2005a}, the ``volume" $\Omega$ in the case of a non-uniform 
trapped Bose gas system can be written $\Omega=\omega_{ho}^{-D}$, where $D$ is the dimensionality 
of the system (in our case $D=3$), and $\omega_{ho}$ the trapping frequency. Further, we note that 
instead of evaluating $P$ only, we have actually evaluated the `pressure' as $P\times \Omega$ which 
has units of energy. Thus, even if $P$ does not have the units of force per unit area, just as 
$\Omega$ does not have the units of volume, the most important fact is that $P \times \Omega$ 
has units of energy. In addition, we show in Appendix \ref{sec:thermo-non-uniform} that our 
Eq.(\ref{eq:sfa-partition-function}) yields the thermodynamic potential for a non-uniform 
trapped Bose gas system in the limit of an infinite number of states and very small 
fluctuations.

\hs The internal energy is given by
\begin{equation}
U\,=\,-\left[ \frac{\partial \ln Q}{\partial \beta}\right]_{\mu,\varphi_F,E} 
\,=\,
\sum_{m=0}^M \left[\frac{\partial q_0(m,T)}
{\partial \beta}\right]_{\mu, \varphi_F, E}, \label{eq:intenergy}
\end{equation}

where the derivative (\ref{eq:intenergy}) has been calculated at constant 
$\mu$, $\varphi_F$, and $E$. Otherwise, if we included the derivatives of $\mu$, 
$\varphi_F$, and $E$ with respect to $\beta$, we find that a physical behavior 
for the thermodynamic properties is not obtainable.

\hs The specific heat is evaluated using 

\begin{equation}
C_v\,=\,\left( \frac{\partial U}{\partial T} \right)_N \,=\, \frac{\partial}
{\partial T} \sum_{m=0}^M \frac{\partial q_o(m,T)}{\partial \beta}. 
\label{eq:specificheat}
\end{equation}

Finally, the entropy is given by

\begin{equation}
S\,=\,\frac{U+P\Omega-\mu N}{T}. \label{eq:exact-entropy}
\end{equation}

In evaluating $S$, we find that $\mu N$ constributes only a small amount to $S$ 
and can therefore be ignored.

\hs The thermodynamic equations (\ref{eq:sfa-pressure}) and (\ref{eq:exact-entropy})
in our paper were previously used for nonuniform gases by Sevincli and Tanatar 
\cite{Sevincli:2007} but in different units than ours. Further, Grether \ea\ \cite{Grether:2003} 
used Eq.(\ref{eq:sfa-pressure}) and Rochin \cite{Rochin:2005a} used (\ref{eq:exact-entropy}) for 
nonuniform Bose gases.

\section{Analysis of the performance of the SFA code}\label{sec:analysis-of-sfa}

\hs In what follows, a brief analysis of the performance of a more advanced 
version of the SFA program than that used in \cite{Al-Sugheir:2001,Al-Sugheir:2002,Ghulam:2007}
is presented. This version has been written in the C programming language and new
features have been included that have enabled us to obtain $\mu(T)$ as in 
Sec.\ref{sec:chemical-potential}, as well as $\varphi_F(m,T)$, and perform a convergence 
check of the iteration procedure.

\subsection{Determining the Chemical Potential}

\begin{figure}[t!]
\begin{center}
\includegraphics[width=8.5cm,bb = 168 455 534 710,clip]{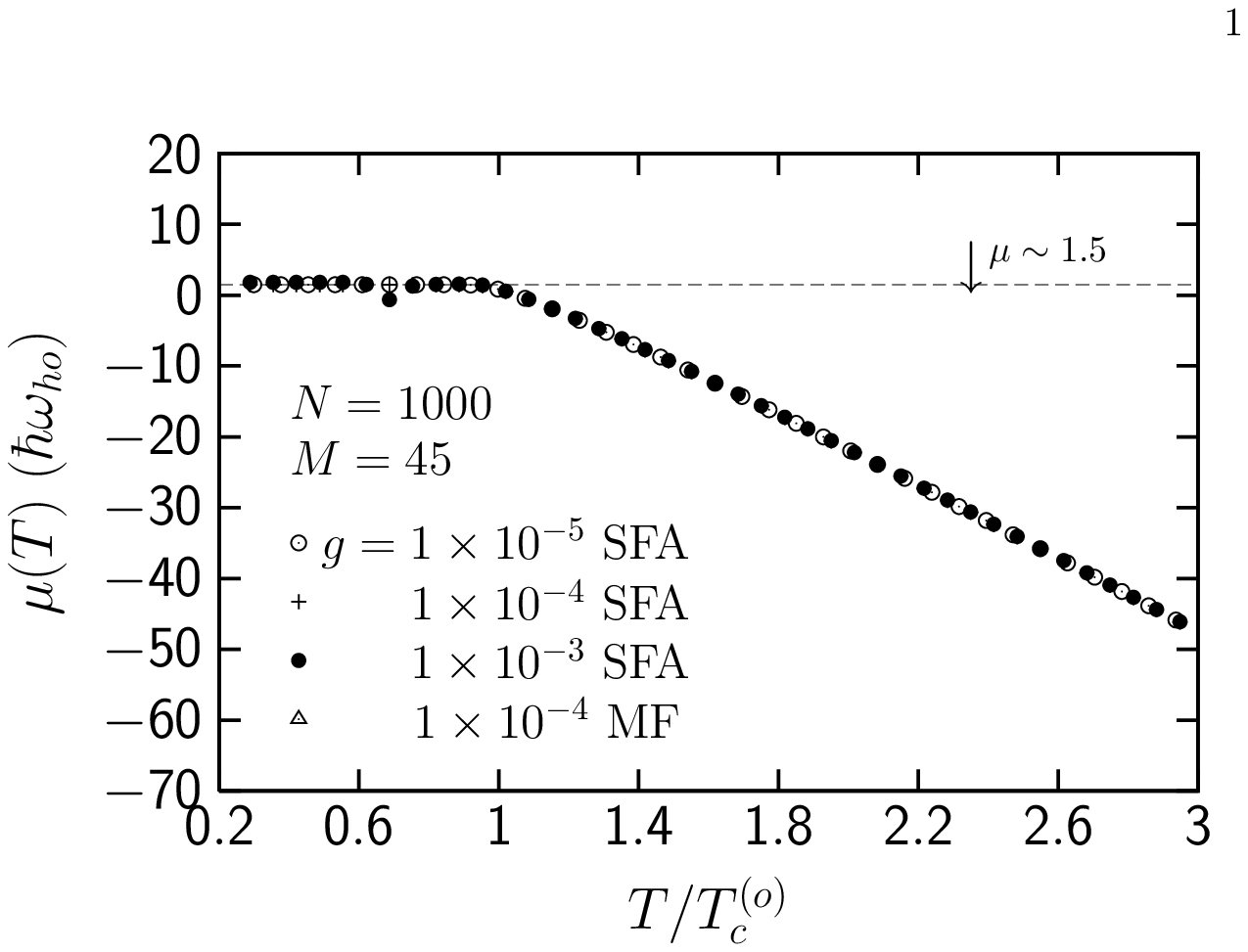}
\caption{The chemical potential $\mu$ for a trapped HS Bose gas of $N=1000$ particles 
as a function of $T$ for different $g$ using SFA and MF techniques. The number of states 
is $M=45$, and the trapping frequency is $\omega_{ho}=2\pi\times 10$ Hz. The dashed
line represents $\mu\sim 1.5$ and $T_c^{(o)}$ is the critical temperature for a noninteracting
Bose gas.} \label{fig:plot.chemp.vs.T.and.g.N1000}
\end{center}
\end{figure}

\hs Figure\ft\ref{fig:plot.chemp.vs.T.and.g.N1000} shows $\mu$ as 
a function of $T/T_c^{(o)}$ for various $g$, where $T_c^{(o)}\,=\,4.5113$ nK has been chosen as our 
reference temperature. These values of $\mu(T)$ have been obtained by the technique of Sec.\ref{sec:chemical-potential}
using $\eta_0(m,T)$. $T_c^{(o)}$ has been evaluated from the noninteracting result 
\cite{Dalfovo:1999}:

\begin{equation}
T_c^{(o)}\,\sim\,0.94\frac{\hbar\omega_{ho}}{k_B}\,N^{1/3},
\label{eq:Tc_and_N} 
\end{equation}
for $N=1000$ and $\omega_{ho}\,=\,2\pi\,\times\,10\,\hbox{Hz}$; this value is used
throughout the rest of the paper. Below the critical temperature, $\mu$ is 
$-$ as expected $-$ equal to the ground-state energy of the harmonic oscillator in 
the weakly-interacting regime, namely, $\sim1.5$ in trap units. Above $T_c^{(o)}$, 
$\mu$ decreases steadily and acquires negative values. The chemical potential seems 
insensitive to the changes in the values of $g$ indicated in the figure. 

\hs For comparison purposes, $\mu(T)$ is also displayed for $g=1\times 10^{-4}$, as obtained by the MF technique in
Sec. \ref{sec:chemical-potential} using $\langle \hat{n}_k(T)\rangle$.
SFA and MF results shown are identical. The fact that $\mu$ below $T_c^{(o)}$ ($\sim 1.5$) 
is close to the ground-state energy indicates that most of the particles reside in the condensate  
at $T<T_c^{(o)}$. These results are in line with those of \"Ohberg and Stenholm \cite{Ohberg:1997} and
Ketterle and van Druten \cite{Ketterle:1996} for the ideal Bose gas. However, when $g$ is increased to order $10^{-2}$, 
$\mu$ below $T_c^{(0)}$ attains values higher than 1.5. 

\begin{figure}[t!]
\begin{center}
\includegraphics[width=8.5cm,bb = 186 455 570 714,clip]{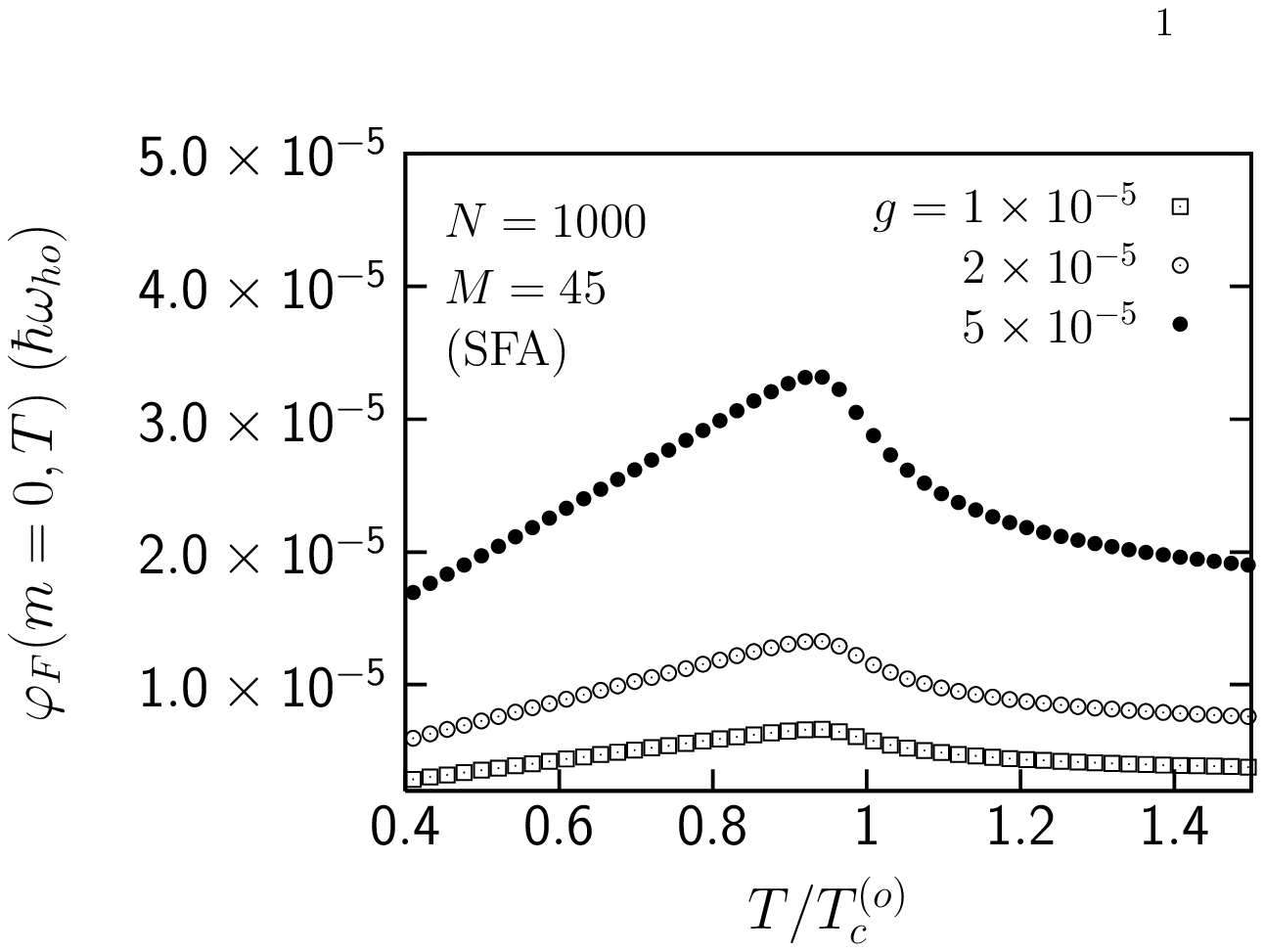}
\caption{SFA fluctuations in the energy $\varphi_F(m=0,T)$ as functions of 
$T$ for a system of $N=1000$ trapped HS bosons and $M=45$ states at three 
interactions $g$. $T_c^{(o)}$ is the critical temperature for a noninteracting Bose gas.}
\label{fig:plot.phi.vs.T.and.g.N1000}
\end{center}
\end{figure}

\begin{figure}[t!]
\begin{center}
\includegraphics[width=8.5cm,bb = 180 458 520 714,clip]{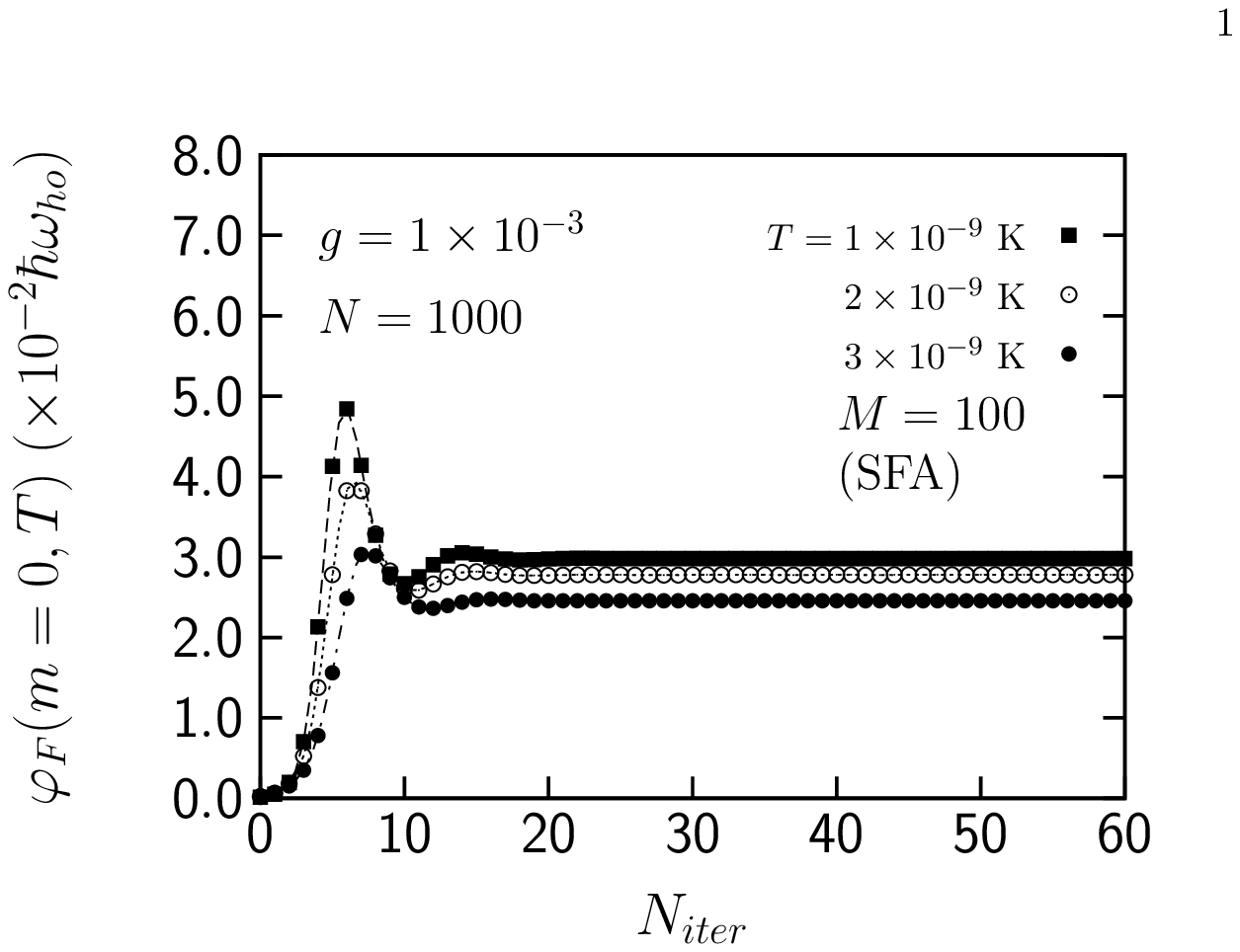}
\caption{SFA fluctuations $\varphi_F(m=0,T)$ as functions of the number of iterations 
$N_{iter}$ for the ground state of a trapped HS Bose gas of $N=1000$ particles and 
$M=100$ states at $g=1\times 10^{-3}$ and different $T$.}
\label{fig:plot.phi.vs.iter.and.T.g1e-3.N1000}
\end{center}
\end{figure}
\begin{figure}[t!]
\begin{center}
\includegraphics[width=8.5cm,bb = 184 458 540 714,clip]{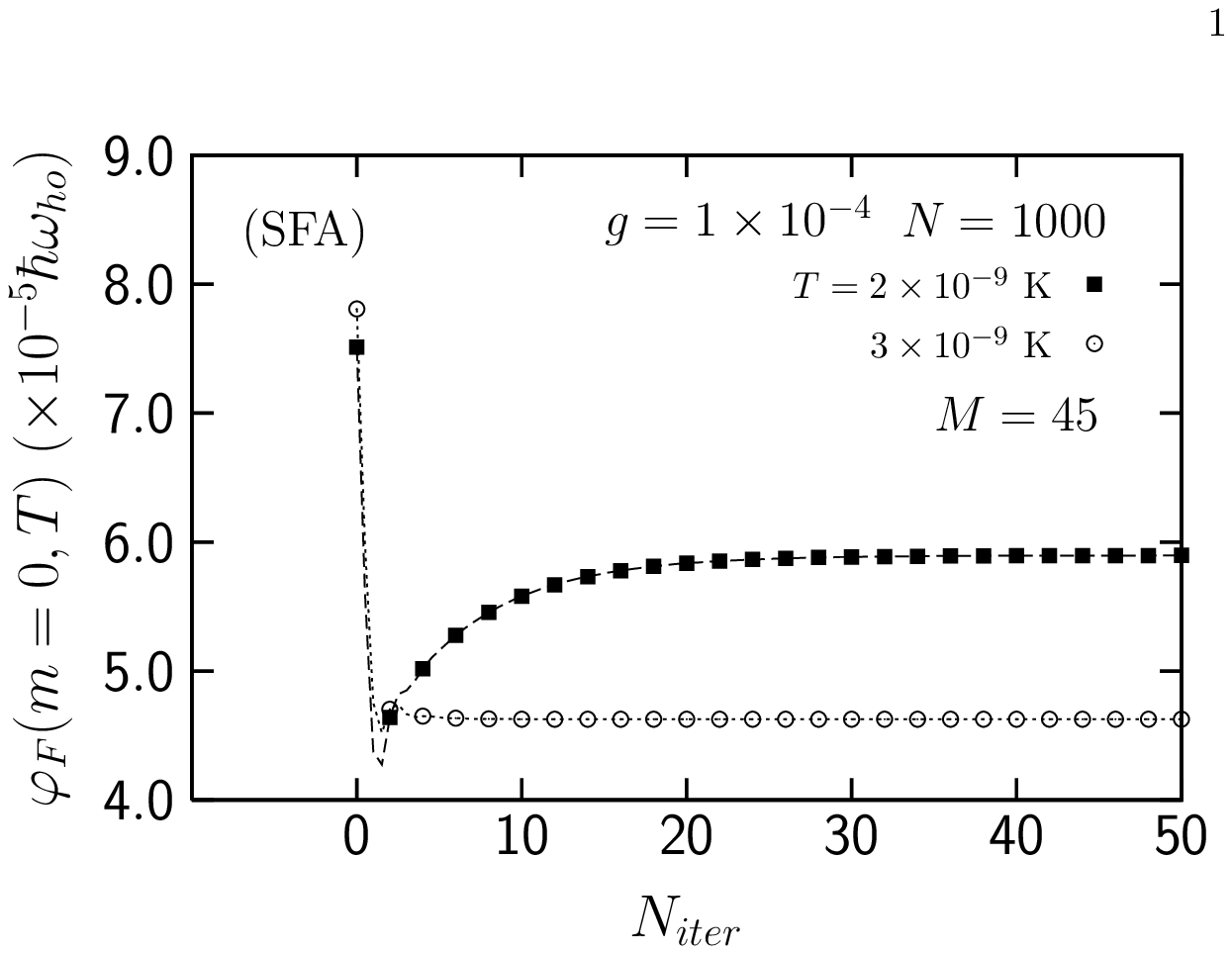}
\caption{SFA fluctuations $\varphi_F(m=0,T)$ as functions of the number of iterations
$N_{iter}$ for the ground state of a trapped HS Bose gas of $N=1000$ particles and $M=45$ states
at $g=1\times 10^{-4}$ and different $T$.} \label{fig:plot.phi.vs.iter.and.T.g1e-4.N1000}
\end{center}
\end{figure}

\begin{figure}[t!]
\begin{center}
\includegraphics[width=8.5cm,bb = 189 456 540 705,clip]{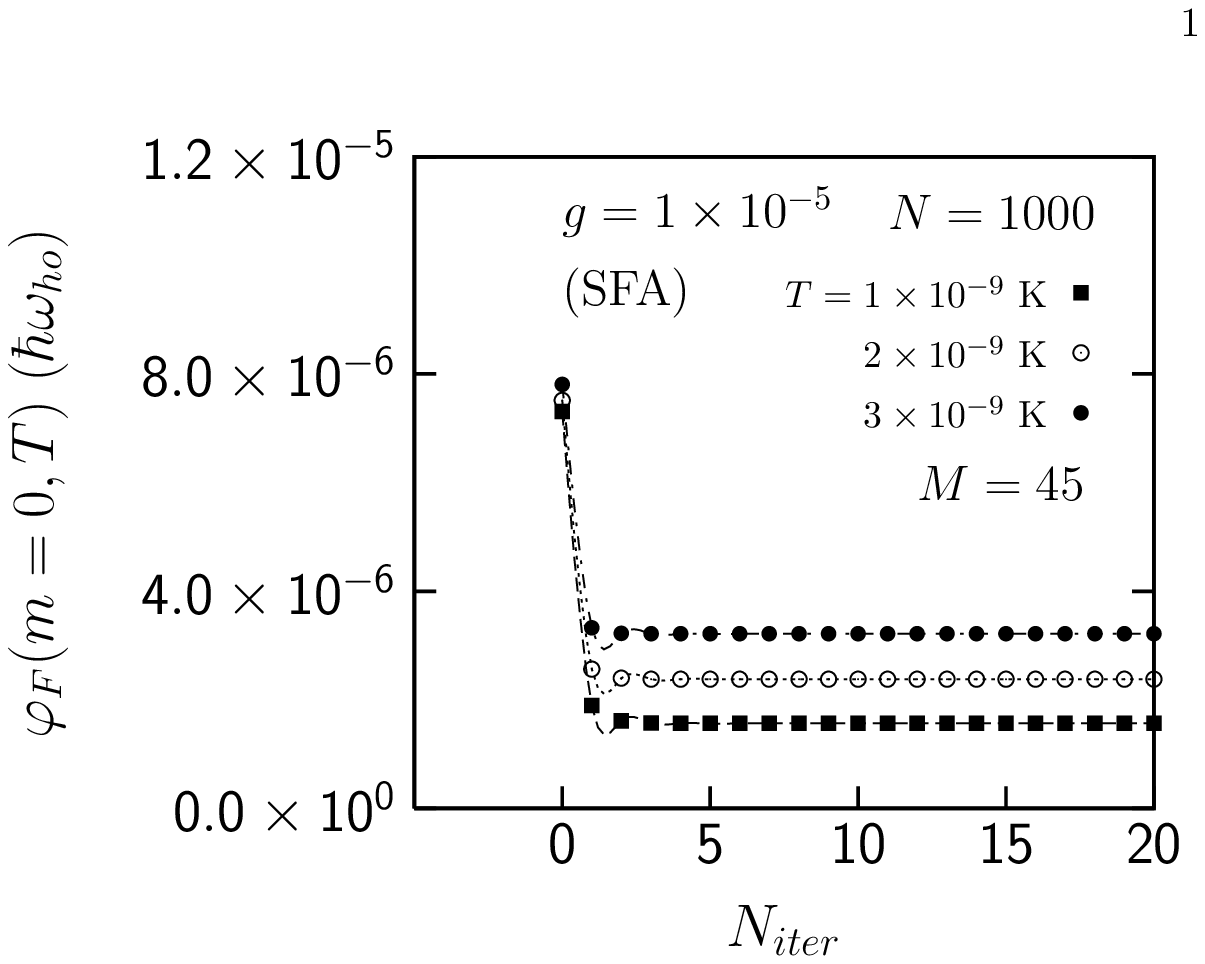}
\caption{SFA fluctuations $\varphi_F(m=0,T)$ as functions of the number of iterations
$N_{iter}$ for the ground state of a trapped HS Bose gas of $N=1000$ particles and $M=45$ states
at $g=1\times 10^{-5}$ and different $T$.} \label{fig:plot.phi.vs.iter.and.T.g1e-5.N1000}
\end{center}
\end{figure}
\begin{figure}[h!]
\begin{center}
\includegraphics[width=8.5cm,bb = 180 459 550 709,clip]{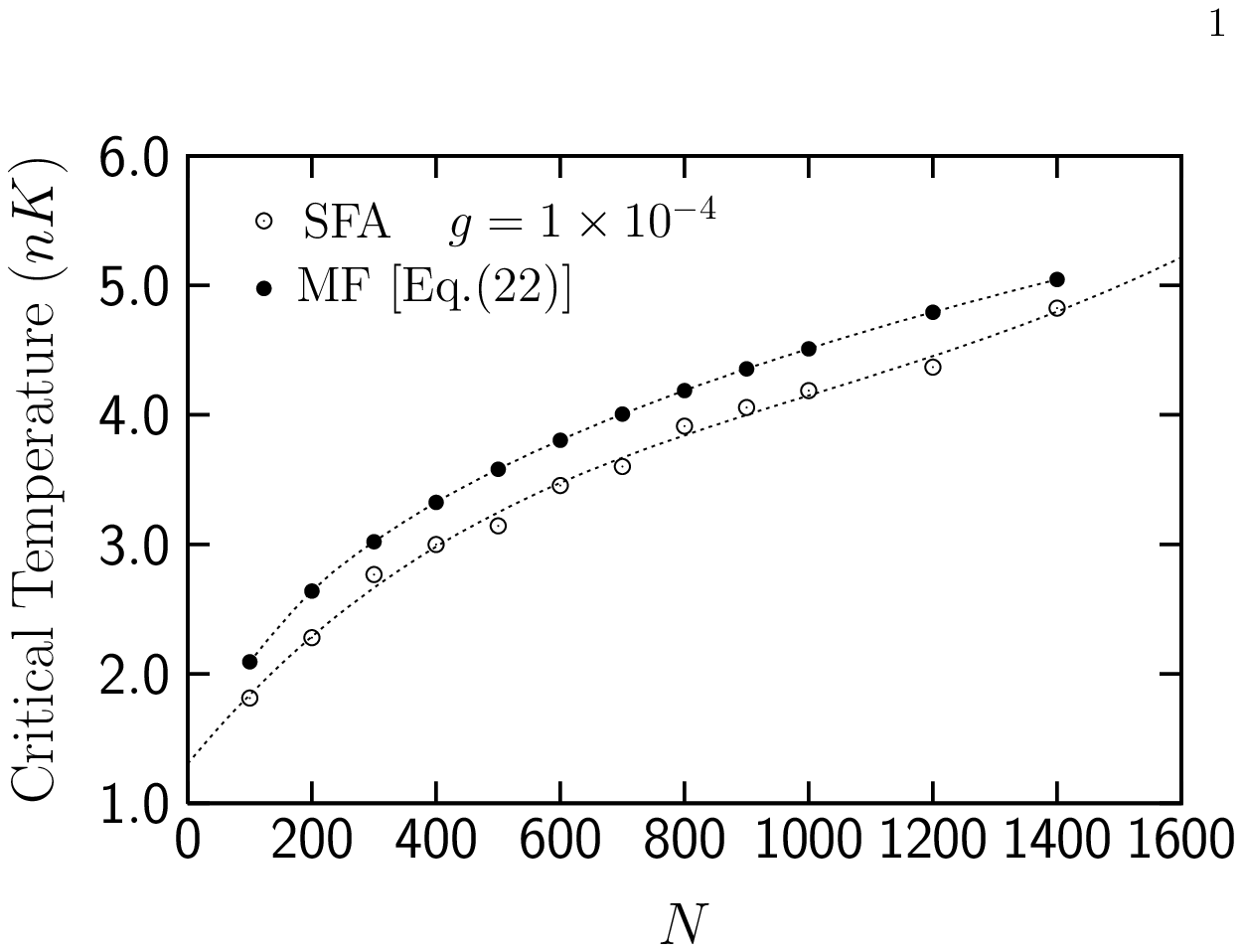}
\caption{Critical temperatures as functions of $N$ for a HS Bose gas in a trap with $g=1\times 10^{-4}$ 
($T_c$; open circles) and for a noninteracting HS Bose gas ($T_c^{(o)}$; solid circles), evaluated 
within SFA and MF approaches, respectively.} \label{fig:plot.TcvsNforSFAandMF.g1e-4}
\end{center}
\end{figure}

\subsection{T-dependence of $\varphi_F$}

\hs Figure\ft\ref{fig:plot.phi.vs.T.and.g.N1000} displays the fluctuations $\varphi_F(m=0,T)$ 
for the ground state ($m=0$) of systems with $g=1 \times 10^{-5}$, $2\times 10^{-5}$, and 
$5 \times 10^{-5}$. The fluctuations increase with $T$ and show a maximum at the 
transition temperature $T_c$, beyond which they are damped. Note that $T_c$ refers to the 
transition temperature in the {\it interacting} 
case; one can infer from Fig.\ft\ref{fig:plot.phi.vs.T.and.g.N1000} that $T_c\,<\,T_c^{(o)}$.

\subsection{Convergence of SFA}

\hs To check the convergence of the iteration procedure outlined in 
Appendix \ref{sec:sfa-iterative-procedure}, $\varphi_F(m,T)$ is plotted versus the 
number of iterations $N_{iter}$ in Figs.\ft\ref{fig:plot.phi.vs.iter.and.T.g1e-3.N1000},
\ft\ref{fig:plot.phi.vs.iter.and.T.g1e-4.N1000}, and \ref{fig:plot.phi.vs.iter.and.T.g1e-5.N1000} for 
several $T$ and $g$. Figures\ft\ref{fig:plot.phi.vs.iter.and.T.g1e-3.N1000} and
\ref{fig:plot.phi.vs.iter.and.T.g1e-4.N1000}, with $g=1\times 10^{-3}$ and $1\times 10^{-4}$, respectively, 
show the need for a large number of iterations ($\sim 25$) for SFA to converge to a steady 
value of $\varphi_F(m=0,T)$. In Fig.\ft\ref{fig:plot.phi.vs.iter.and.T.g1e-5.N1000}, with $g=1\times 10^{-5}$, 
on the other hand, $\varphi_F(m,T)$ converges rapidly after only 4 iterations for all the temperatures indicated. 
Thus, the higher the interaction, the larger the number of iterations needed to achieve convergence. It is, 
therefore, of the utmost importance to check the convergence of SFA calculations.

\subsection{Critical temperature}

\hs Figure\ft\ref{fig:plot.TcvsNforSFAandMF.g1e-4} shows a comparison between the critical temperature for an
interacting (using SFA) and a noninteracting (using MF) system as a function of $N$. The solid circles pertain to $T_c^{(o)}$
of Eq.\ft(\ref{eq:Tc_and_N}) for a noninteracting, and the open circles to $T_c$ for an interacting, system with $g=1\times 10^{-4}$.
The SFA results have been obtained from the peak positions of the specific heat capacities.

\hs A considerable difference is noted between the two results, which shows that the critical 
temperature drops with increasing $g$. Further, the trend in $T_c$ is almost the same as $T_c^{(o)}$.
The critical temperature $T_c$ obtained by SFA increases with $N$, as expected on physical grounds.

\subsection{Failure of SFA at higher $g$}\label{sec:failure-of-sfa}

\hs It is found that SFA breaks down at $g>1\times 10^{-3}$. The reason for this is
that, at the higher interactions, the fluctuations become very large. As a result the exponent 
$\langle \hat{E}_m \rangle\,-\,\mu\,-\,\varphi_m$ becomes negative in

\begin{eqnarray}
&&\eta_0(m,T)\,\equiv\,\frac{1}{2}\left\{\frac{1}{\exp[\beta(\langle \hat{E}_m
\rangle\,-\,\mu\, +\varphi_m)]\,-\,1}\,+\,
\frac{1}{\exp[\beta(\langle \hat{E}_m \rangle\,-\,\mu\,-\,\varphi_m)]\,-\,1}\right\}. \nonumber \\
\end{eqnarray}

This causes the second term on the right-hand side of $\eta_0(m,T)$ to acquire a negative sign as well. 
If a large number of states is negative, the sum of $\eta_0(m,T)$, weighted by the degeneracies $d_m$, may 
also turn out to be negative; whereas it is supposed to yield the total number of particles for an optimal 
$\mu$. Essentially this kills the calculation, since $f(\mu,T)$ [Eq.(\ref{eq:minimization-function})] will 
be far away from the true minimum. We may call this the `sign problem' in the SFA method.

\section{Results and Discussion}\label{sec:results-and-discussion}

\hs In this section the interrelationship between $N$, $M$, and $g$, the energy as a 
function of $T$, the occupancy of states, and the thermodynamic properties are presented.

\subsection{Interrelationship between $N$, $M$, and $g$}

\begin{figure}[t!]
\begin{center}
\includegraphics[width=8.5cm,bb = 195 525 549 773,clip]{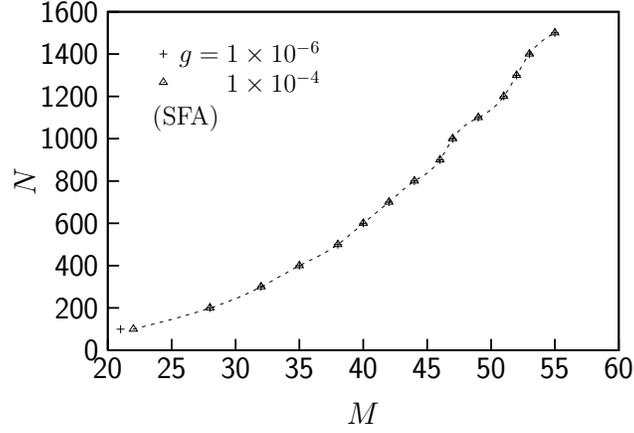}
\caption{Interrelationship between the number of particles $N$ and the number of states $M$
of trapped HS Bose gases at two values of $g$. The points $(N,\,M)$ pertain to the classical
limit $3Nk_B$ of the specific heat capacity.} \label{fig:NvsM}
\end{center}
\end{figure}

\hs Figure \ref{fig:NvsM} displays the dependence of $N$ on $M$ for two values of $g$. For each
$N$, the value of $M$ is adjusted to the classical limit of the specific heat capacity, $C_v=3Nk_B$, 
in the weakly-interacting regime. One can see that, for a given $g$, an increase in $N$ necessitates
a rise in $M$; although the rise in $M$ is fairly small for each increment of $N$, namely
100. Generally speaking, then, one would not need an infinite number of states to accommodate numbers of
particles of order 100. It is when one seeks systems with $N\sim 10^6$ that the number
of states is expected to shoot up to very large values. It is further found that the change of $M$ with
$N$ is independent of $g$ in the weakly-interacting regime, and $M$ is mainly determined by $N$.

\subsection{Energy fluctuations}

\begin{figure}[t!]
\begin{center}
\includegraphics[width=8.5cm,bb = 179 521 549 775,clip]{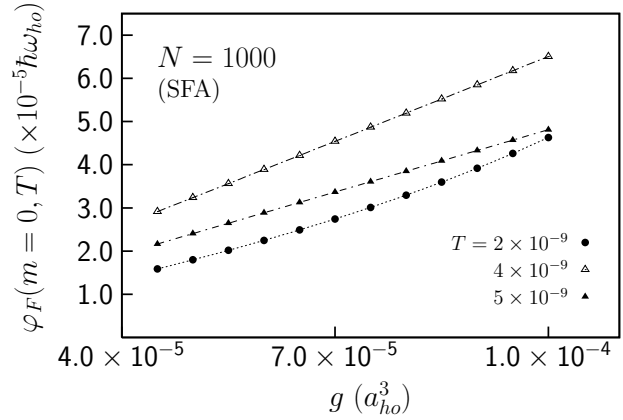}
\caption{SFA ground state fluctuation $\varphi_F(m=0,T)$ vs $g$ at several $T$ for $N=1000$ particles.} 
\label{fig:phi-vs-g}
\end{center}
\end{figure}

\hs Figure \ref{fig:phi-vs-g} displays $\varphi_F(m=0,T)$ versus $g$ for $N=1000$ at various $T$.
$\varphi_F(m=0,T)$ rises almost linearly as $g$ is increased. Further, $\varphi_F(m=0,T)$ rises with
$T$ up to the transition temperature beyond which it drops.

\subsection{Energy}

\begin{figure}[t!]
\begin{center}
\includegraphics[width=8.5cm,bb = 164 454 534 708,clip]{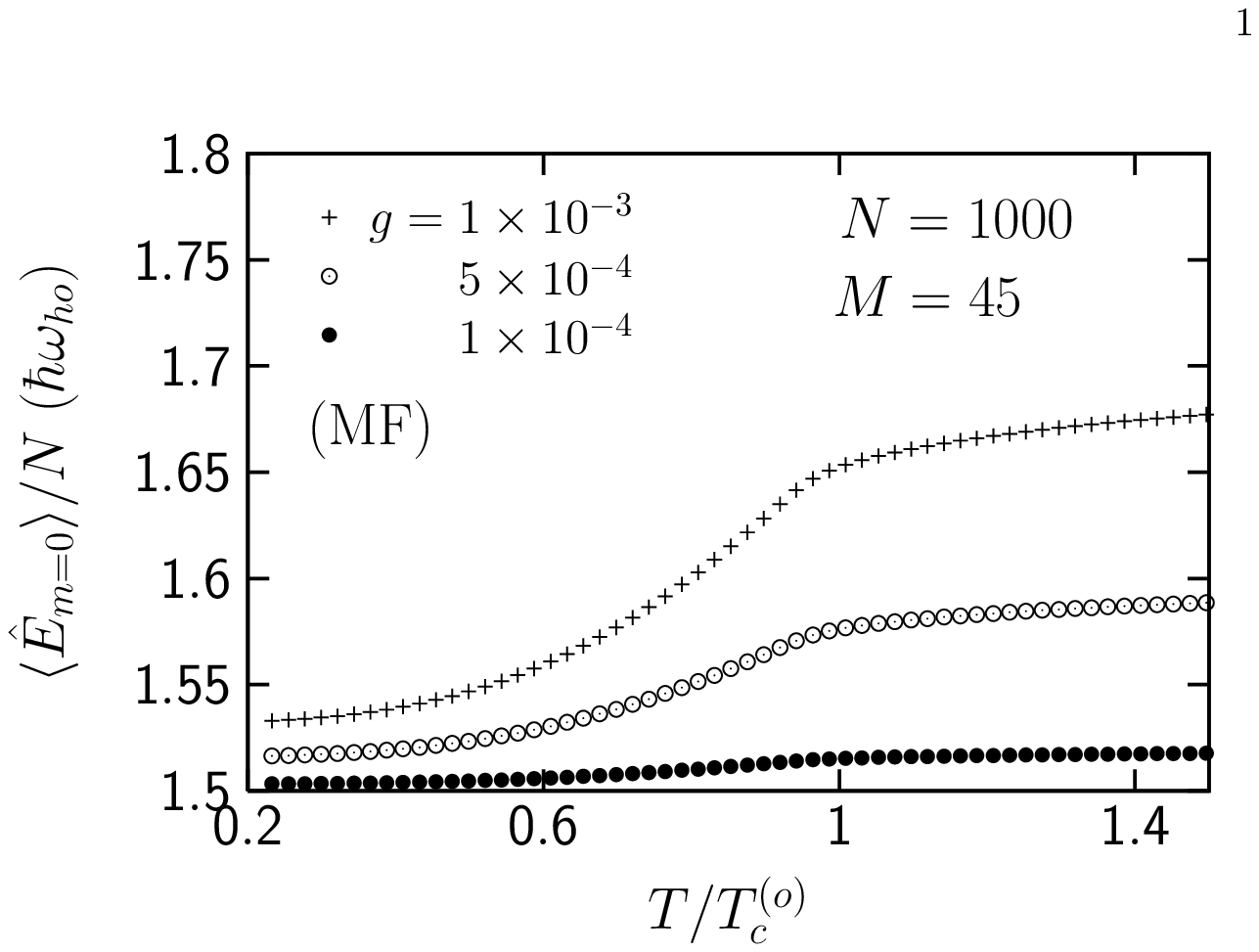}
\caption{The MF energy per particle for $m=0$ as a function of $T$ for 
$N=1000$ trapped HS bosons, $M=45$ states and three interaction parameters $g$.} 
\label{fig:plot.E.vs.T.and.g.N1000}
\end{center}
\end{figure}

\hs Figure\ft\ref{fig:plot.E.vs.T.and.g.N1000} displays the MF energy per particle
$\langle \hat{E}_{m=0} \rangle/N$ in the condensate state, as obtained from Eq.\,(\ref{eq:avenergy_m}), 
for $N=1000$ particles and various $g$. The energy rises initially at a faster rate below $T_c$ than above it.
As expected, the energy also rises with $g$. However, even though $g$ and the changes in it are small,
the variations in $E$ vs $T$ are relatively considerable; for example, at $T/T_c^{(0)}=1.5$, one observes
$\Delta E>0.05$. This indicates that a small change in $g$ causes $\langle \hat{n}_k(T)\rangle$ to 
attain a noticeably stronger $T$-dependence, i.e., $\langle \hat{n}_k(T)\rangle$ is very sensitive to
variations in $g$. As the temperature rises, more particles are thermally excited out of the condensate to higher HO states. 
As a result, the energy arising from the interaction of the condensate with higher excited states increases 
with $T$, causing an increase in the total energy.

\begin{figure}[t!]
\begin{center}
\includegraphics[width=8.5cm,bb = 168 454 534 708,clip]{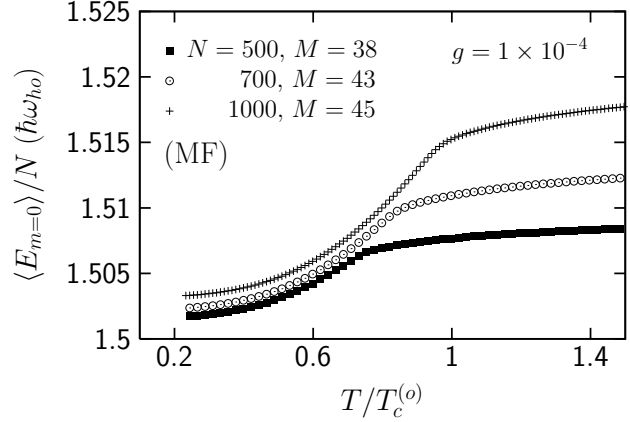}
\caption{The MF energy per particle for a fixed $g=1\times 10^{-4}$ and different $M$ and $N$ of a 
trapped HS Bose gas. In each case, $M$ has been adjusted so as to attain a reasonable value for the 
classical limit of the specific heat capacity.} \label{fig:plot.E.vs.T.and.N.g1e-4}
\end{center}
\end{figure}

\begin{figure}[t!]
\begin{center}
\includegraphics[width=8.5cm,bb = 178 515 537 780,clip]{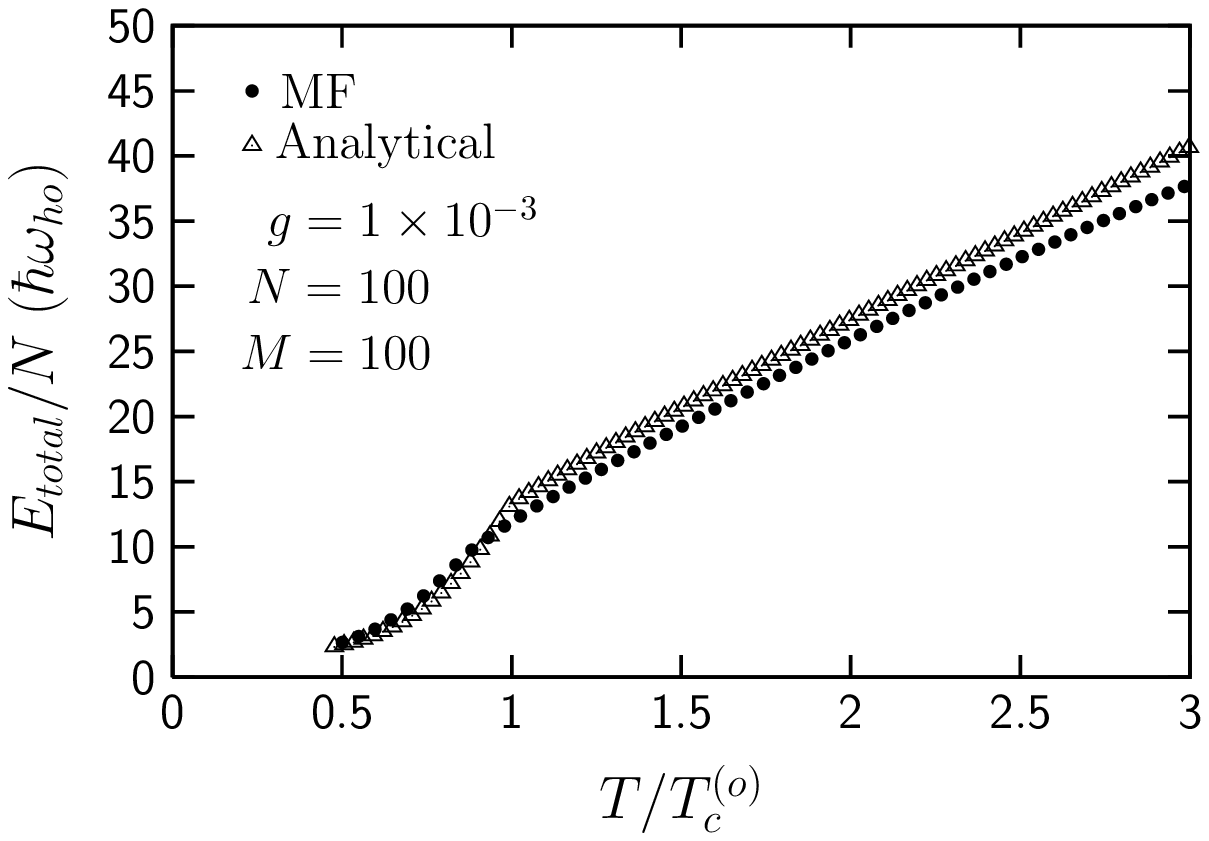}
\caption{The total MF energy per particle $E_{total}/N$ for an interacting trapped HS Bose gas of $N=100$
particles, $M=100$ states, and $g=1\times 10^{-3}$. The open triangles show the analytical results 
by Su \ea\ $^{35}$ and the solid circles the MF results of the present work.}
\label{fig:plot.E.comparisons.w.analytical.N100M100g1e-3}
\end{center}
\end{figure}

\begin{figure}[t!]
\begin{center}
\includegraphics[width=8.5cm,bb = 174 519 537 775,clip]{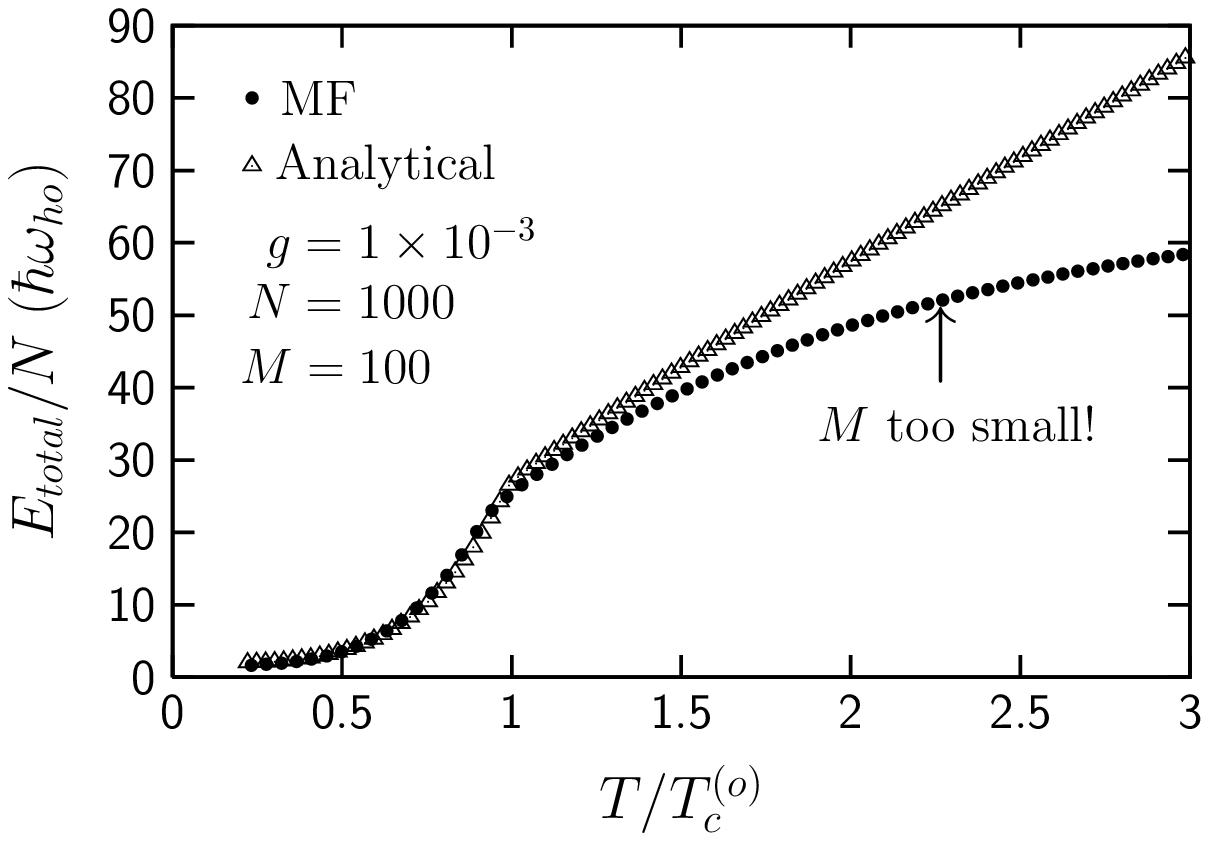}
\caption{The total MF energy per particle $E_{total}/N$ for an interacting trapped HS Bose gas of $N=1000$
particles, $M=100$ states, and $g=1\times 10^{-3}$. The open triangles show the analytical results by 
Su \ea\ $^{35}$ and the solid circles the MF results of the present work.}
\label{fig:plot.E.comparisons.w.analytical.N1000M100g1e-3}
\end{center}
\end{figure}

\hs Figure\ft\ref{fig:plot.E.vs.T.and.N.g1e-4} displays the behavior of $\langle \hat{E}_{m=0} \rangle/N$ 
when $g$ is fixed and $N$ is varied. Clearly, the energy rises with $N$; as $N$ increases, 
more particles occupy higher HO states. Notice that a relatively considerable change in $N$ is required
in order to increase $E$ vs. $T$. For example, one needs $\Delta N=300$ to get $\Delta E=5\times 10^{-3}$
at $T/T_c^{(o)}=1.5$. This shows that a change in $N$ has a weaker effect on the energy than a change in $g$.

\subsection{Comparison to analytical results}

\hs The {\it total} mean-field energy of this work is

\begin{equation}
E_{total}\,=\,\sum_{m=0}^M \langle \hat{E}_m \rangle \langle \hat{n}_m \rangle,
\label{eq:MFtotalE}
\end{equation}

where $\langle \hat{E}_m \rangle$ is given by Eq.\ft(\ref{eq:avenergy_m}).
Figure \ref{fig:plot.E.comparisons.w.analytical.N100M100g1e-3} displays a comparison between the MF $E_{total}/N$ 
(Eq.\ft\ref{eq:MFtotalE}) for $N=100$, $g=1\times 10^{-3}$, and $M=100$, and the analytical $E_{total}/N$ 
as obtained by Eqs.(47) and (67) of Su \ea\ \cite{Su:2006} for the same parameters $N$ and $g$. The two results almost match for 
$T<T_c^{(o)}$, and good agreement is obtained for $T>T_c^{(o)}$.
Figure \ref{fig:plot.E.comparisons.w.analytical.N1000M100g1e-3} is the same as
Fig.\ft\ref{fig:plot.E.comparisons.w.analytical.N100M100g1e-3}; but for $N=1000$. Again, the two results almost
match for $T<T_c^{(o)}$; but beyond that, the deviation from the analytical results is substantial, particularly
above $T\,=\,1.5T_c^{(o)}$. The reason for this large deviation goes back to the 
limited maximum number of states that we can use. Had we been able to apply $M>100$, the MF result in 
Fig.\ref{fig:plot.E.comparisons.w.analytical.N1000M100g1e-3} would have approached the analytical result as $M$
was increased since the excited particles would have been able to occupy higher states. 
In Fig.\ft\ref{fig:plot.E.comparisons.w.analytical.N100M100g1e-3}, the number of states was sufficient to 
accommodate all the excitations, and this is why there is good agreement above $T_c^{(o)}$. Further, the good
agreement below $T_c^{(o)}$ arises from the fact that, at $T<T_c^{(o)}$, all particles are in the $(m=0)$ 
condensate state. Thus, there is no need for a large number of states to describe the system below $T_c^{(o)}$. 
In other words, the effect of excitations becomes significant only beyond $T=T_c^{(o)}$.

\hs Thus, the MF model developed in Sec.\ref{sec:method} works well. Likewise, the minimization technique 
of Sec.\ref{sec:chemical-potential} for evaluating the chemical potential is quite effective. 
A supercomputer should enable one to reach states much higher than $M=100$, particularly for the 3D case.

\begin{figure}[t!]
\begin{center}
\includegraphics[width=8.5cm,bb = 167 457 534 708,clip]{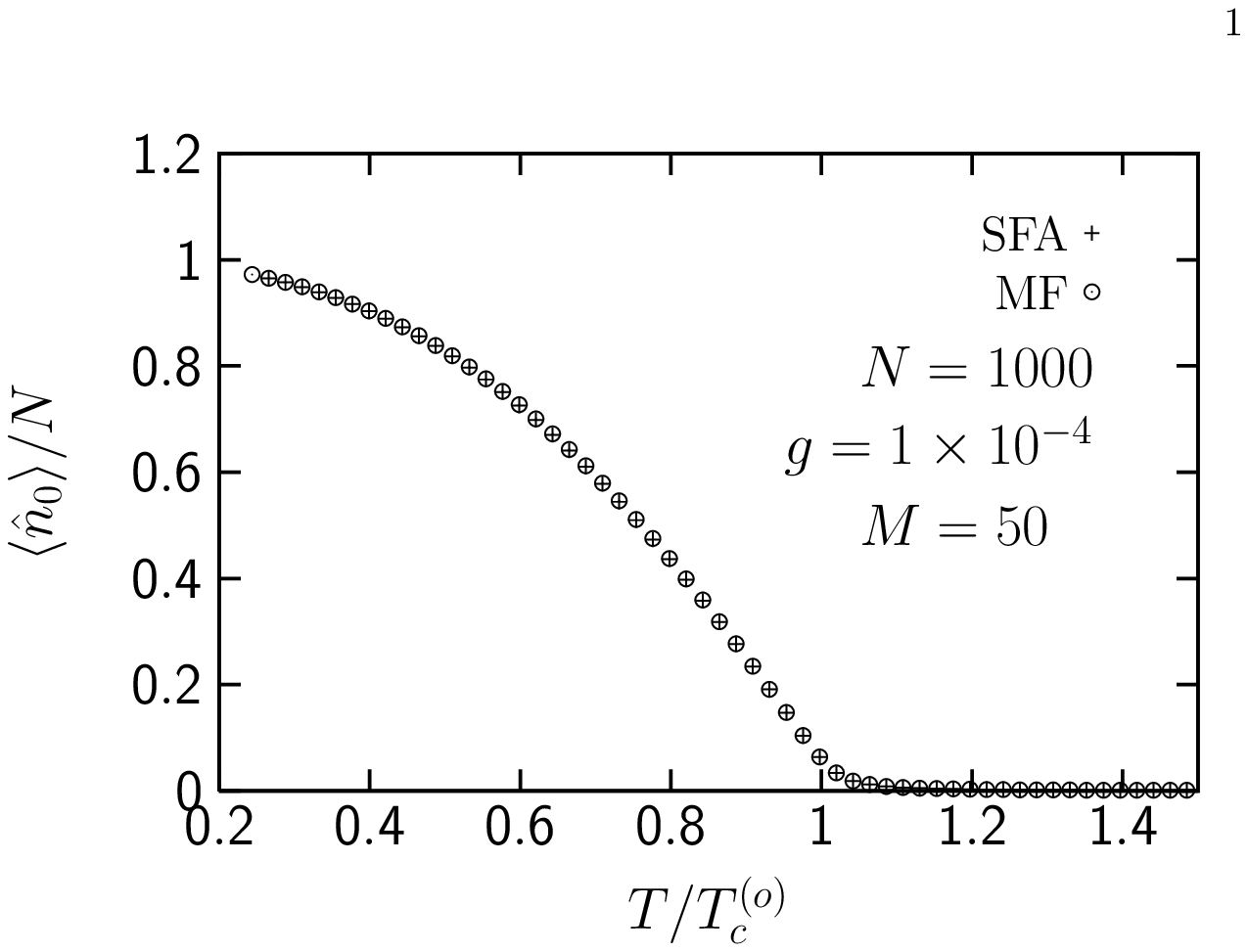}
\caption{SFA (crosses) and MF (open circles) fractional occupation numbers $\langle \hat{n}_0/N\rangle$ as 
functions of $T$ for the HO ground state, $m=0$, of a system of $N=1000$ trapped HS bosons, $M=50$ states,
with $g=1\times 10^{-4}$.} \label{fig:plot.n0.MFandSFA.N1000.g1e-4}
\end{center}
\end{figure}

\subsection{Occupancy of states}\label{sec:occupancy-of-states}

\begin{figure}[t!]
\begin{center}
\includegraphics[width=8.5cm,bb = 167 457 534 708,clip]{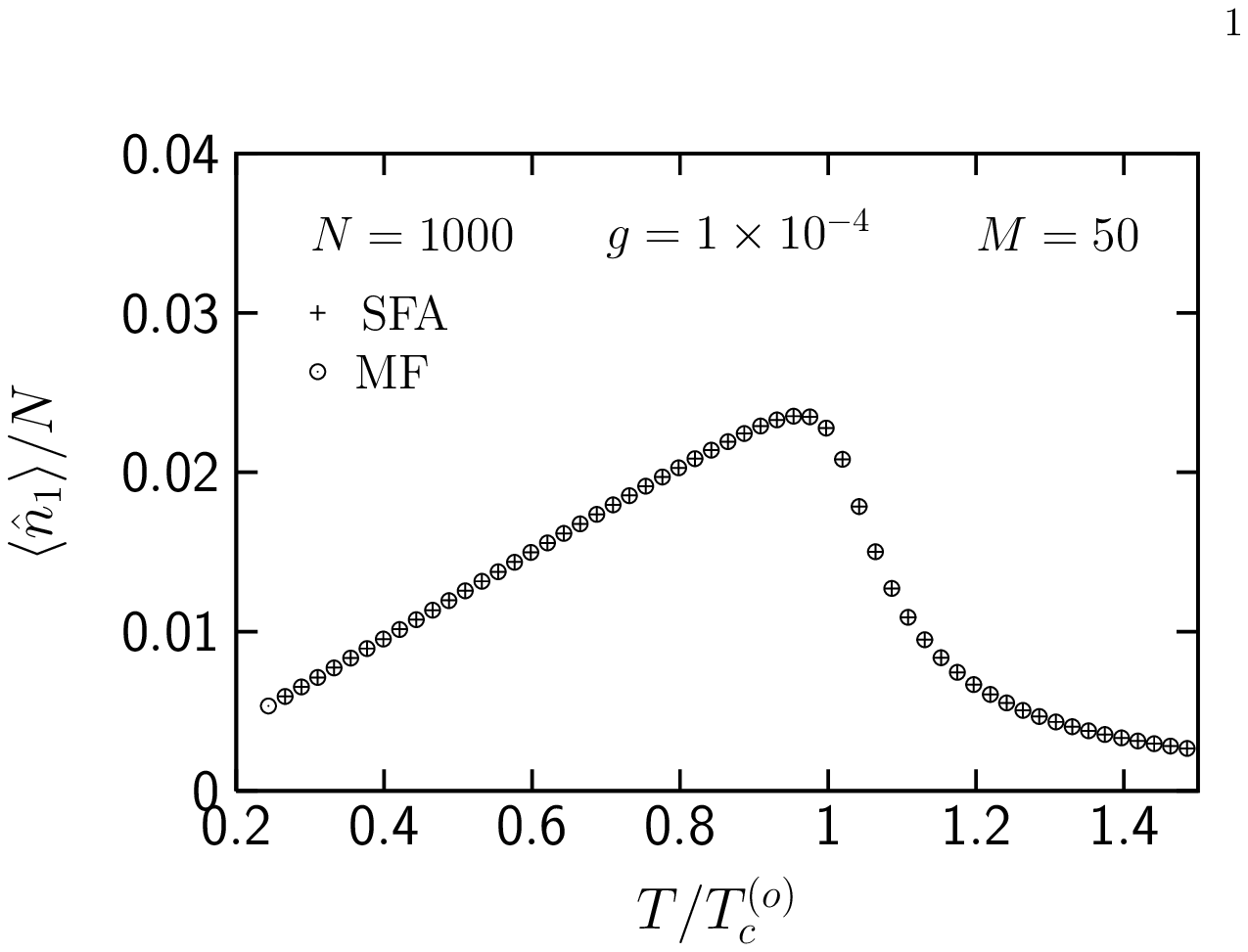}
\caption{SFA (crosses) and MF (open circles) fractional occupation numbers $\langle \hat{n}_1/N\rangle$ as 
functions of $T$ for the HO ground state, $m=1$, of a system of $N=1000$ trapped HS bosons, $M=50$ states,
with $g=1\times 10^{-4}$.} \label{fig:plot.n1.MFandSFA.N1000.g1e-4}
\end{center}
\end{figure}

\begin{figure}[t!]
\begin{center}
\includegraphics[width=8.5cm,bb = 167 457 538 710,clip]{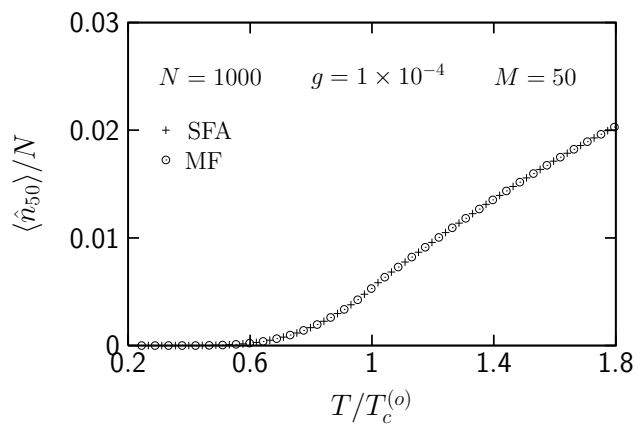}
\caption{SFA (crosses) and MF (open circles) fractional occupation numbers $\langle \hat{n}_{50}/N\rangle$ as 
functions of $T$ for the HO ground state, $m=50$, of a system of $N=1000$ trapped HS bosons, $M=50$ states,
with $g=1\times 10^{-4}$.} \label{fig:plot.n50.MFandSFA.N1000.g1e-4}
\end{center}
\end{figure}

\hs Figure\ft\ref{fig:plot.n0.MFandSFA.N1000.g1e-4} shows the fractional 
occupation number (condensate fraction) $\langle \hat{n}_0 \rangle/N$ versus  
$T$ of bosons in the HO ground state, evaluated using SFA [crosses, $\eta_0(m=0,T)$, Eq.\,(\ref{eq:eta0})] 
and MF [open circles, $\langle \hat{n}_{k =0} \rangle$, Eq.\,(\ref{eq:Bose-Einstein.occup})] 
for a system of $N=1000$ particles, $M=50$ states, and $g=1\times 10^{-4}$. The SFA  
$\eta_0(m=0,T)$ is identical to the MF $\langle \hat{n}_{m=0}(T) \rangle$ because the fluctuations in 
the energy $\varphi_F$ are much smaller than the energy itself at  
$g=1\times 10^{-4}$. One observes that, slightly above $T_c^{(o)}$, 
$\langle \hat{n}_0 \rangle/N$ drops to zero. In Fig.\ft\ref{fig:plot.n1.MFandSFA.N1000.g1e-4} we display 
for the same system the fractional occupation number for the first excited HO state 
$\langle \hat{n}_1 \rangle/N$ versus $T$. One can see that, while $\langle \hat{n}_0 \rangle/N$ 
decreases from 1 to almost zero at $T_c^{(0)}$, $\langle \hat{n}_1 \rangle/N$ increases steadily to a maximum at 
$T\approx 0.9T_c^{(0)}$ above which it drops systematically to zero. This 
kind of behavior has also been observed by Ketterle \ea\ \cite{Ketterle:1996}. 
Going further, we present in Fig.\ft\ref{fig:plot.n50.MFandSFA.N1000.g1e-4} 
$\langle \hat{n}_{50} \rangle/N$ for the same system. The occupation 
number in this rather high state keeps rising with $T$. If $M=50$ had not been the highest 
state assigned in this particular calculation, $\langle \hat{n}_{50} \rangle$ would have reached a maximum beyond which it would have declined to 
zero, thus allowing other particles to be excited to states higher than $m=50$ at higher 
$T$. Hence, in general, the maximum of $\langle \hat{n}_m \rangle$ is expected to shift 
to higher $T$ as $m$ becomes higher. It turns out, then, that $M$ chiefly controls the overall distribution of 
$N$ particles among the discrete states. It is to be emphasized that when one restricts
$M$ (to, e.g., 50), the BE occupation number at any given state $m$ along with the chemical potential 
will be overestimated. This is because the bosons are shuffled back to lower states and are redistributed 
among them, particularly for $g>0.01$. In this case, any states higher than 
$m=50$ remain practically unoccupied. 

\hs Figure\ft\ref{fig:plot.n1.MFandSFA.N1000.g1e-4} shows that, when 
$T$ reaches $T_c^{(0)}$, about 2.2$\%$ of the particles occupy the first excited state; and Fig.\ref{fig:plot.n50.MFandSFA.N1000.g1e-4} shows that only 
$\sim 0.5\%$ occupy the state $m=50$. But as $T$ rises further above $T_c^{(0)}$, the particles leave the 
condensate state $m=0$ and occupy very high states such as $m=50$; for example, at $T/T_c^{(0)}=1.8$ we 
see that $\langle \hat{n}_{50} \rangle/N\sim 0.02$, i.e., 2$\%$ of the particles occupy our highest state.
Nevertheless, for the values of $g$ used in this investigation
($g\le 1\times 10^{-3}$), $M=50$ seems to be a reasonable value, since $\langle \hat{n}_{50} \rangle/N$ is only
2$\%$ in Fig.\ft\ref{fig:plot.n50.MFandSFA.N1000.g1e-4}. 

\hs Further, the maximum values of $N$ and $g$ we are able to use, namely $N=1000$ and $g=1\times 10^{-3}$,
correspond to \begin{math}Na_s\,=\,N g/(4\,\pi)\,=\,7.96\times 10^{-2}\end{math}, which is much less than 1.
Consequently, we are way below the Thomas-Fermi regime defined by $Na_s\gg 1$.

\subsection{Thermodynamic properties}

\subsubsection{Critical Temperature}\label{sec:critical_temperature}

\begin{figure}[t!]
\begin{center}
\includegraphics[width=8.5cm,bb= 167 454 533 709,clip]{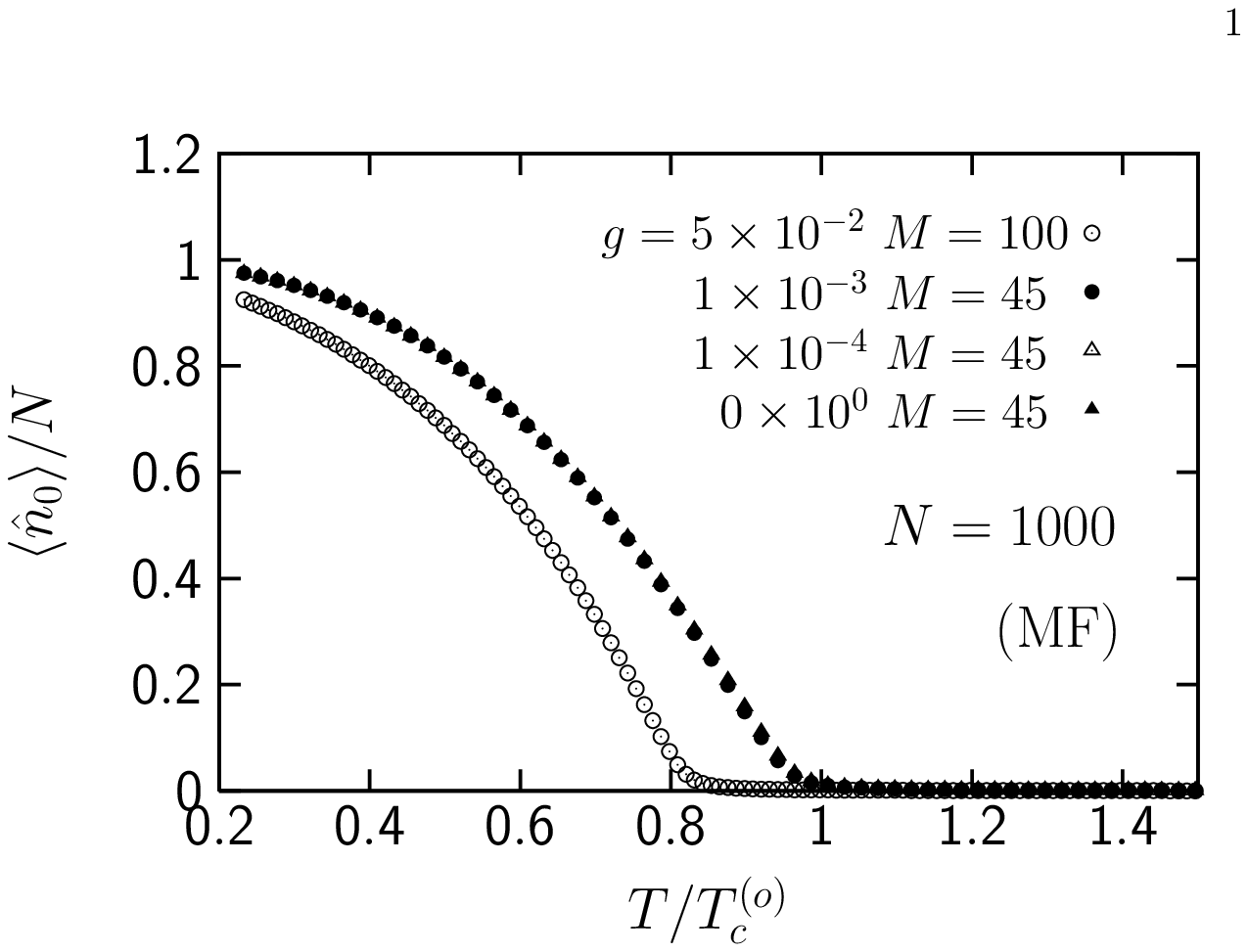}
\caption{The MF fractional occupation number $\langle \hat{n}_0 \rangle/N$ for $N=1000$ trapped HS bosons
and $M=100$ states, for $g=0.00$ (noninteracting system) and $g=0.05$ (interacting system).}
\label{fit:plot.n0.for.g.N1000}
\end{center}
\end{figure}

\hs Figure\ft\ref{fit:plot.n0.for.g.N1000} displays $\langle \hat{n}_0 \rangle/N$ for a fixed $N=1000$ 
particles and different $g$. $T_c$ does not change appreciably between $g=1\times 10^{-4}$ and 
$1 \times 10^{-3}$; but on increasing $g$ to 0.05, $T_c$ drops noticeably by $\sim -0.2T_c^{(o)}$ which 
is about 10 times the result of Giorgini \ea\ \cite{Giorgini:1996}: $\Delta T\,\sim\,-0.016T_c^{(o)}$, 
the shift in $T_c^{(o)}$ being given by these authors as

\begin{equation}
\frac{\delta T_c}{T_c^{(0)}}\,=\,-1.3 \frac{a}{a_{ho}} N^{1/6}.
\label{eq:Giorginis-shift-of-Tc}
\end{equation}

Thus, there is a large discrepancy between the shift of $T_c$ as obtained from our MF approach and that 
from Eq.\ft(\ref{eq:Giorginis-shift-of-Tc}). The reason for this large discrepancy is
that, at higher interactions, the current HO wave function $-$originally designed for a weakly-interacting system$-$
becomes inadequate for the description of a trapped Bose gas with $g=0.05$. This is because the bosonic wave functions
begin to broaden substantially away from their width described by $\exp(-x^2/2)$ in Eq.(\ref{eq:HOwavefunction}), and one 
should rather apply a parameterized Gaussian with a variable width described by $\exp(-\alpha x^2)$, where $\alpha$ is
an adjustable parameter obtainable by variational techniques. We thus anticipate that, if an optimized wave function
is used in our calculations, $T_c$ will approach that of Giorgini \ea\ \cite{Giorgini:1996}.
We will investigate this issue in the future.

\subsubsection{Internal energy}

\begin{figure}[t!]
\begin{center}
\includegraphics[width=8.5cm,bb = 176 457 540 710,clip]{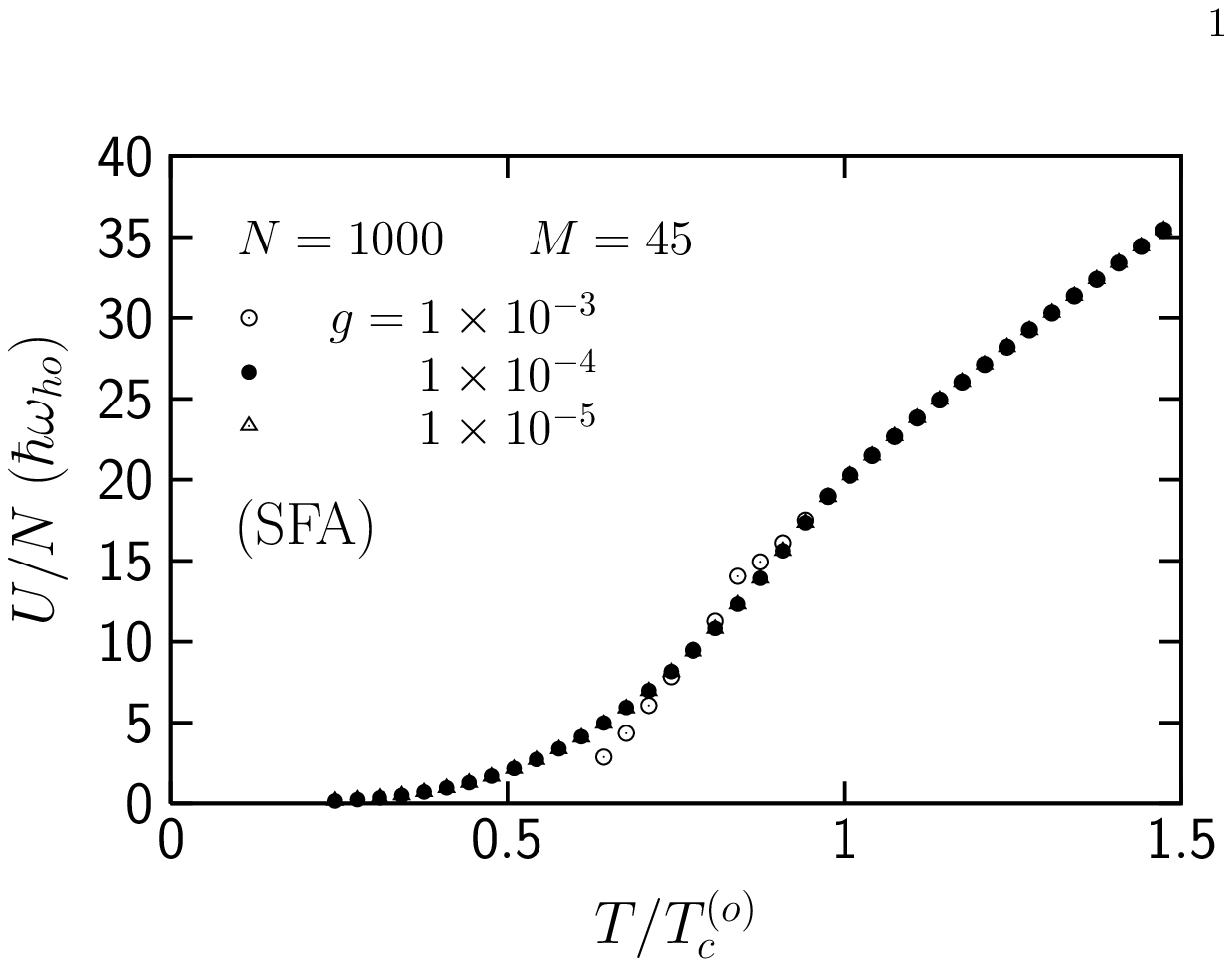}
\caption{The SFA internal energy per particle $U/N$ as a function of $T$ for
$N=1000$ trapped HS bosons, $M=45$ states, and different $g$.} \label{fig:plot.U.vs.g.N1000}
\end{center}
\end{figure}

\begin{figure}[t!]
\begin{center}
\includegraphics[width=8.5cm,bb = 176 457 540 710,clip]{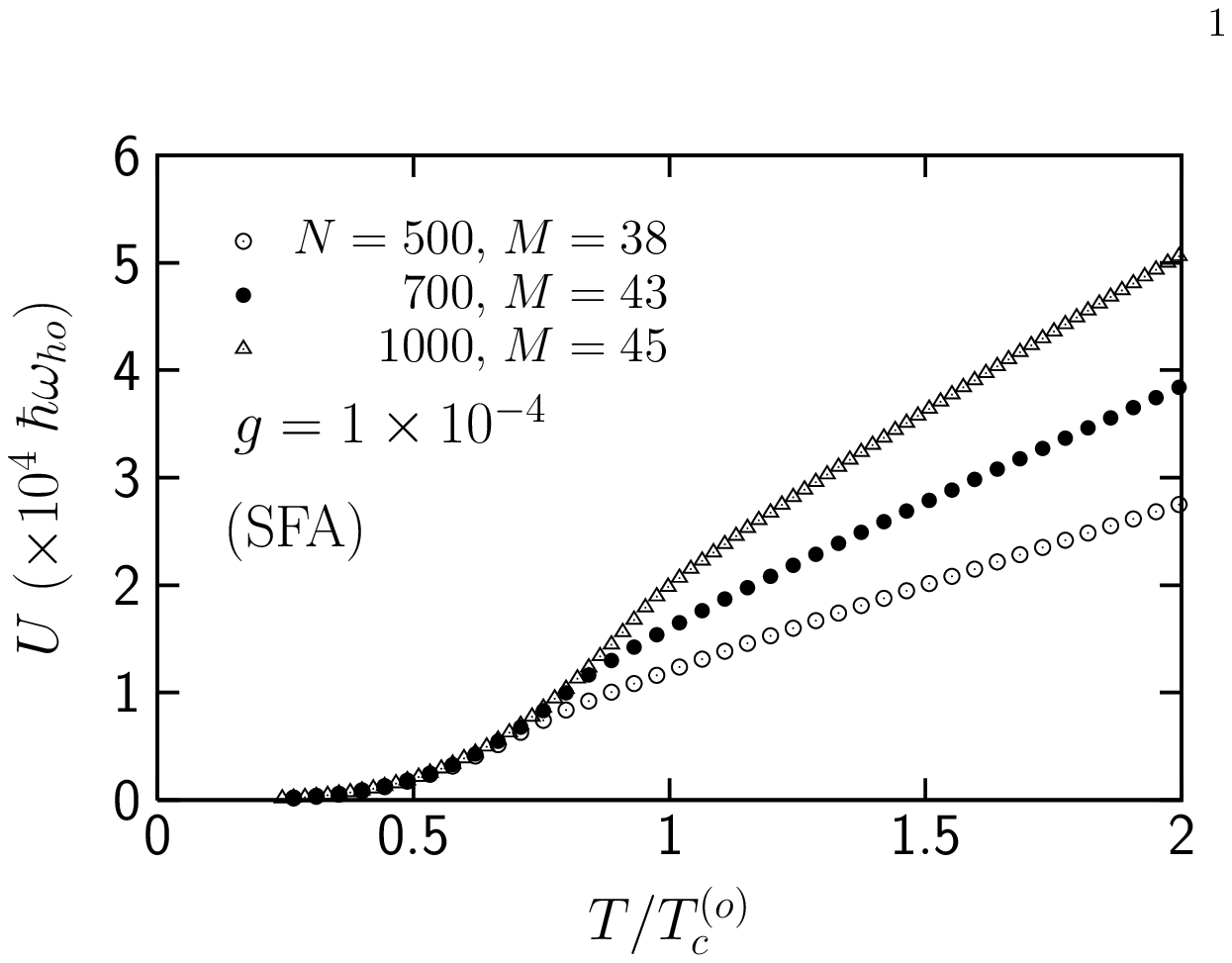}
\caption{The SFA internal energy per particle $U/N$ a a function of $T$ for a fixed $g=1\times 10^{-4}$, and different $N$ and $M$ of trapped HS Bose gases.} \label{fig:plot.U.vs.N.g1e-4}
\end{center}
\end{figure}

\hs Figure\ft\ref{fig:plot.U.vs.g.N1000} displays the internal energy per particle $U/N$ as
a function of $T$ for different $g$, keeping $N=1000$ fixed. $U$ rises with $T$; but it does 
not change appreciably with the variations in $g$. This indicates that the rate of two-body 
collisions is low in the weakly-interacting regime. The rise of $U$ with $T$ is partly due to 
the thermal expansion of the Bose gas towards the edges of the external trapping potential. 
As it approaches the edges of the trap, the gas gains more potential energy. 
Figure\ft\ref{fig:plot.U.vs.N.g1e-4} shows the variations of $U(T)$ with $N$, $g$ being kept fixed 
at $1\times 10^{-4}$. A change in $N$ causes a much more pronounced change in the thermodynamic 
properties than a change in $g$ within the displayed range. This is because, as $N$ rises, the 
density of the system rises, and therefore the rate of collisions increases, altering the thermodynamic properties
substantially. In this case, $U(T)$ is initially insensitive to the changes in $g$ and $N$ at 
lower $T$ up to $T\sim 0.75T_c^{(0)}$; but then towards $T=T_c^{(o)}$ the deviations in the 
behavior of $U(T)$ become noticeable at various $N$.

\subsubsection{Specific heat}\label{sec:Cv-properties}

\begin{figure}[t!]
\begin{center}
\includegraphics[width=8.5cm,bb = 175 456 535 710,clip]{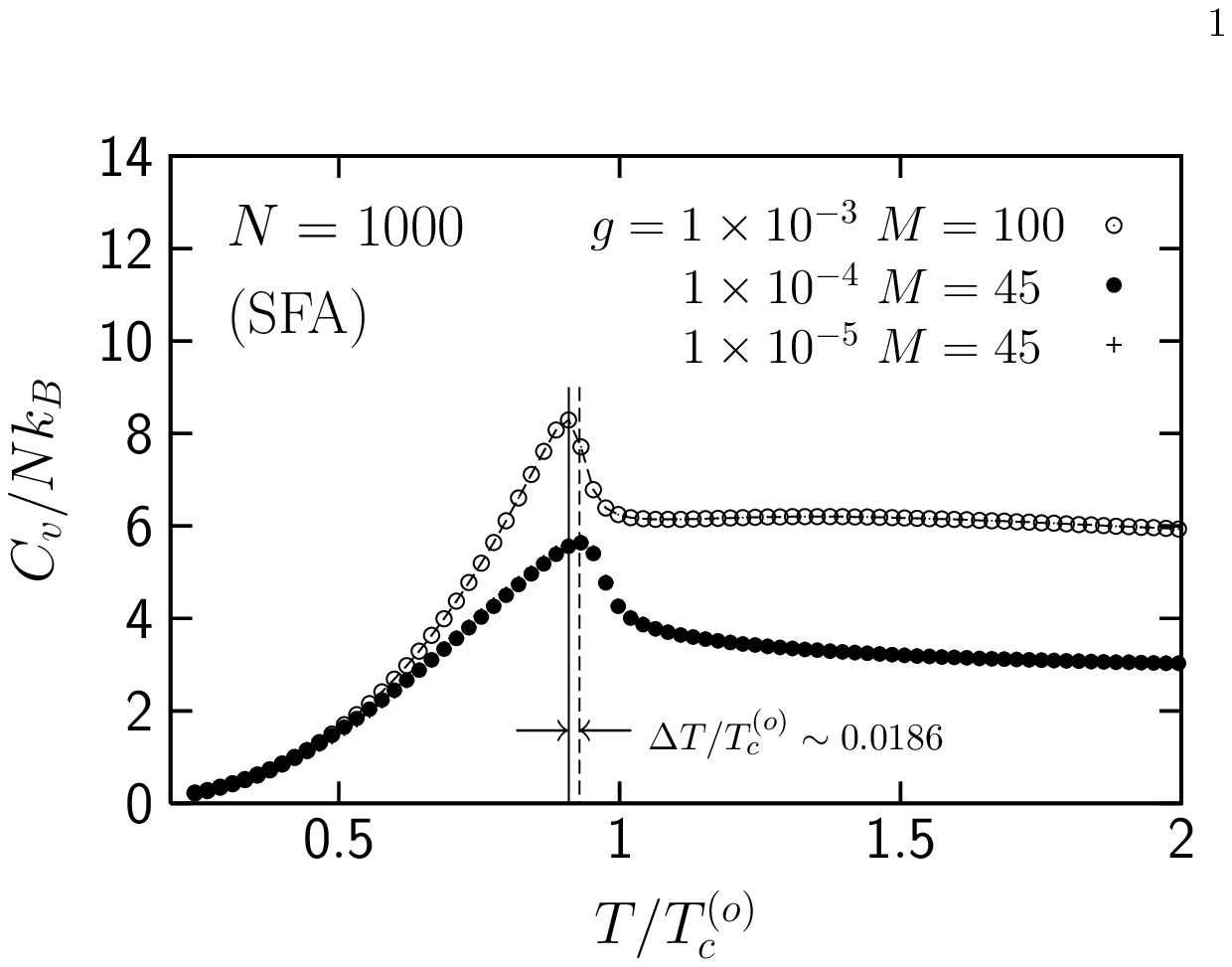}
\caption{The SFA specific heat capacity per particle $C_v/N$ in units of $k_B$ for
$N=1000$ trapped HS bosons for different $g$ and $M$. The shift in the transition 
temperature is indicated.} \label{fig:plot.Cv.vs.g.N1000}
\end{center}
\end{figure}

\hs Figure\ft\ref{fig:plot.Cv.vs.g.N1000} shows the specific heat capacity per 
particle $C_v/N$ versus $T$ in units of $k_B$ for a system of $N=1000$ bosons 
and different $g$. The behavior of $C_v\,=\,(\partial U/\partial T)_\Omega$ 
conforms with the behavior of $U$ for the same systems in Fig.\ft\ref{fig:plot.U.vs.g.N1000} 
at $g=1\times 10^{-5}$ and $1\times 10^{-4}$. Initially, the slope of $U$ increases 
with $T$ up to $T_c$, after which it begins to decline and then stabilizes 
at higher $T$ towards the classical regime. The appearance of a peak in $C_v$ can 
also be directly attributed to the behavior of $\mu(T)$ in the neighborhood of $T_c$ displayed in 
Fig.\ref{fig:plot.chemp.vs.T.and.g.N1000} since $C_v$ given by Eq.(\ref{eq:specificheat}) is connected 
to $\mu(T)$ via $q_0(m,T)$ [Eq.(\ref{eq:q_0(m,T)})]. Note that below $T_c$, $\partial \mu/\partial T \sim 0$, and
when it approaches $T_c^{(o)}$, $\partial \mu/\partial T$ suddenly drops to a negative
value (see Fig.\ft\ref{fig:plot.chemp.vs.T.and.g.N1000}). This abrupt change
in $\mu(T)$ and $\partial \mu/\partial T$ causes a jump in $q_0(m,T)$ and $\partial q_0(m,T)/\partial T$ which
explains the appearance of a peak in $C_v$. 
Alternatively, this `bump' in $C_v$ is clearly
an 'order-disorder' transition since $C_v\,=\,T (\partial S/\partial T)_\Omega$. We note that 
$C_v$ stabilizes at approximately the classical value of $\sim 3N k_B$ \cite{Pethick:2002} for 
$g=1\times 10^{-5}$ and $1\times 10^{-4}$. However, for $g=1\times 10^{-3}$, a larger 
number of states $M=100$ was taken to accommodate the need of 
particles to occupy higher HO states. This leads to a larger classical limit
for $C_v$, namely, $6Nk_B$, as we are now approaching the strongly-interacting regime. 
In fact, it was found that, for large interactions $g\ge 1\times 10^{-3}$, SFA calculations do not converge for 
$M < 100$. Figure\ft\ref{fig:plot.Cv.vs.g.N1000} shows also that the signature 
of the transition $-$i.e., the peak of $C_v$ $-$ shifts to lower $T$ by an 
amount $\Delta T/T_c^{(o)}\sim -0.0186$ as $g$ rises from $1\times 10^{-4}$ to $1\times 10^{-3}$. 
Using formula (\ref{eq:Giorginis-shift-of-Tc}), we get for $\Delta T/T_c^{(o)}$ between $g\,=\,4\pi a_c\,=\,1\times 
10^{-3}$ and $1\times 10^{-4}$ the value $\Delta T/T_c^{(o)}\sim -2.944\times 10^{-4}$ K
which is two orders of magnitude smaller than our SFA result for $\Delta T$. 

\begin{figure}[t!]
\begin{center}
\includegraphics[width=8.5cm,bb = 178 457 535 710,clip]{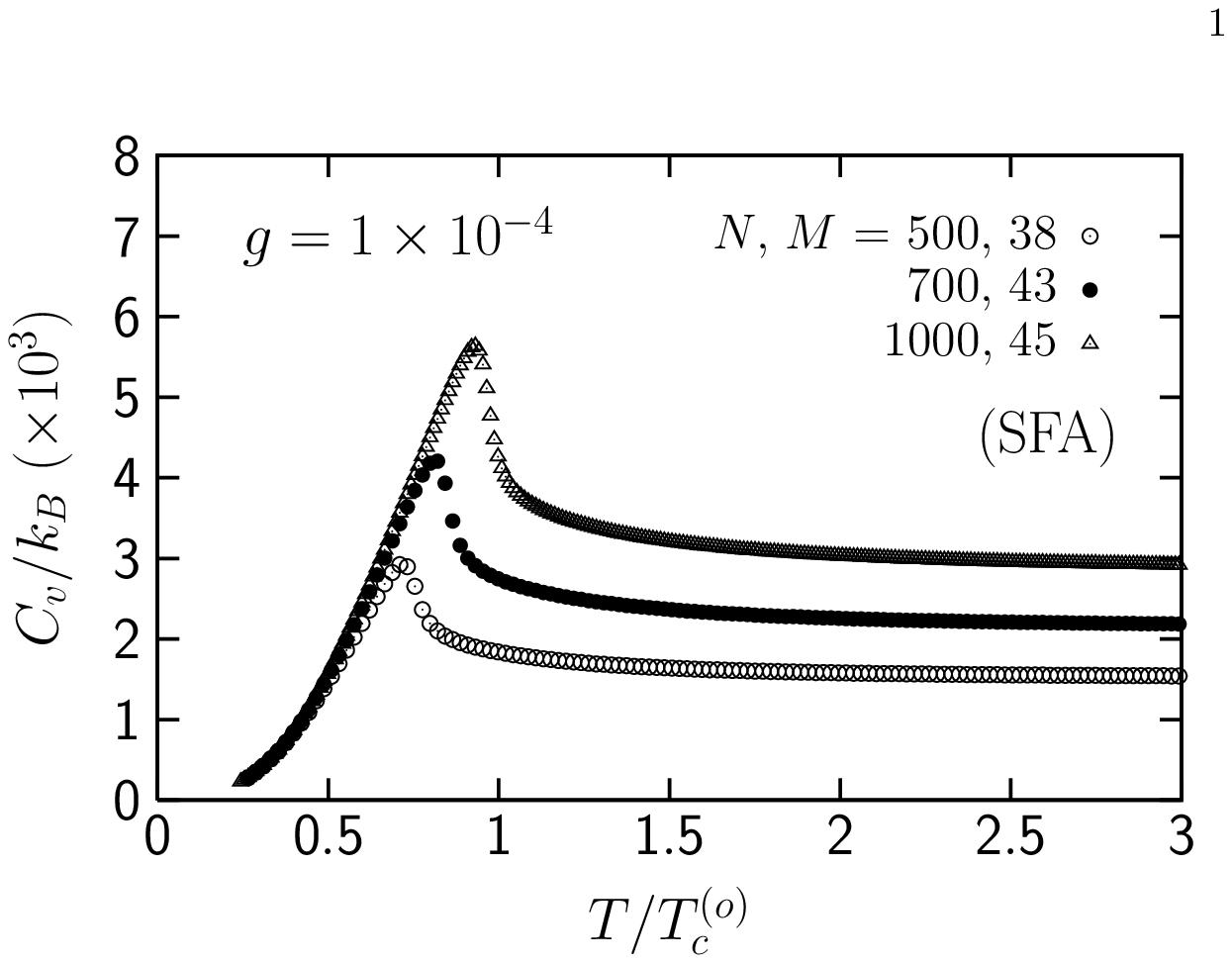}
\caption{The SFA specific heat capacity per particle $C_v/N$ in units of $k_B$ at
a fixed $g=1\times 10^{-4}$ and different $N$ and $M$ of trapped HS Bose gases.}
\label{fig:plot.Cv.vs.N.g1e-4}
\end{center}
\end{figure}

\hs Figure\ft\ref{fig:plot.Cv.vs.N.g1e-4} shows $C_v$ for a fixed $g$ and various $N$. 
We observe that the amplitude of the peak in $C_v$ rises with increasing $N$. Further,
the peak shifts to higher $T$ as $N$ is increased, 
indicating that $T_c$ increases. This is in line with the 
result for dilute Bose gases $T_c^{(0)}\sim 0.94(\hbar\omega_{ho}/k_B) N^{1/3}$ 
\cite{Dalfovo:1999}.

\subsubsection{Entropy}

\begin{figure}[t!]
\begin{center}
\includegraphics[width=8.5cm,bb = 168 456 535 710,clip]{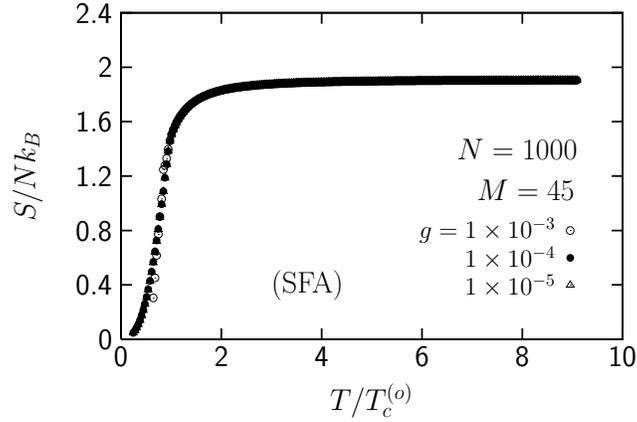}
\caption{The SFA entropy per particle $S/N$ in units of $k_B$ as a function of $T$ for $N=1000$ trapped
HS bosons, $M=45$ states, and different $g$.} \label{fig:plot.S.vs.g.N1000}
\end{center}
\end{figure}

\begin{figure}[t!]
\begin{center}
\includegraphics[width=8.5cm,bb = 168 456 535 710,clip]{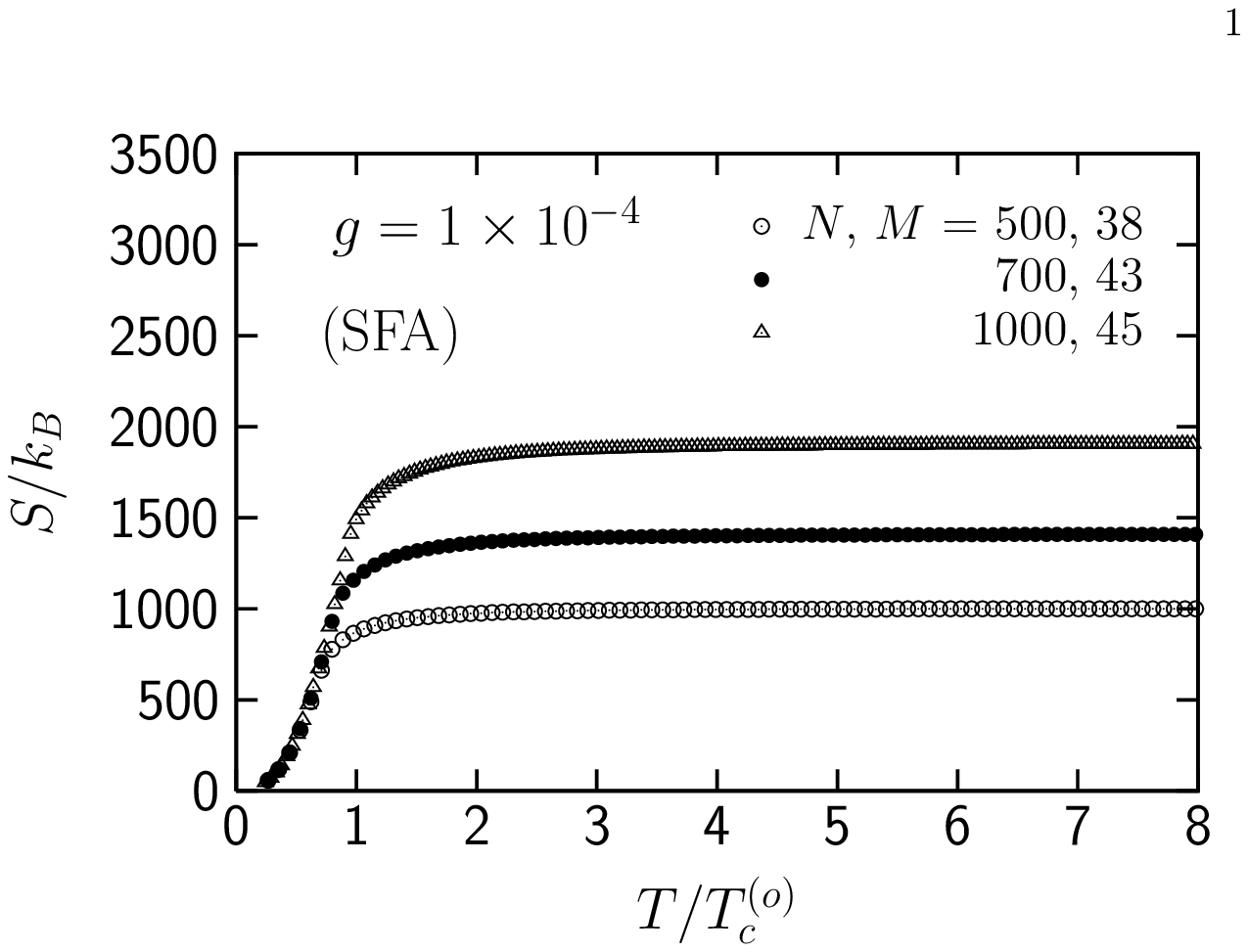}
\caption{The SFA entropy as a function of $T$ in units of $k_B$ for a fixed $g=1\times 10^{-4}$ and
different $N$ and $M$ of trapped HS Bose gases.} \label{fig:plot.S.vs.N.g1e-4}
\end{center}
\end{figure}

\hs Figure\ft\ref{fig:plot.S.vs.g.N1000} displays the entropy per particle $S/N$ in units of $k_B$ 
versus $T$ for different $g$ at a fixed $N=1000$. The entropy rises steeply below $T_c^{(0)}$, but 
then kinks over towards a plateau. Figure\ft\ref{fig:plot.S.vs.N.g1e-4} is the same as 
Fig.\ft\ref{fig:plot.S.vs.g.N1000}, except that we keep $g$ fixed at $1\times 10^{-4}$ and vary $N$. 
Thus, as for $C_v$, $S$ eventually reaches the classical limit at higher $T$. Again the change 
in $g$ from $1\times 10^{-5}$ to $1\times 10^{-3}$ does not have an appreciable effect 
on the thermodynamic properties. The entropy reveals higher order as $T\rightarrow 0$; in
fact, as $T\rightarrow T_c^{(o)}$ the previously mentioned order-disorder transition is observed.

\subsubsection{Pressure}

\begin{figure}[h!]
\begin{center}
\includegraphics[width=8.5cm,bb = 165 456 538 710,clip]{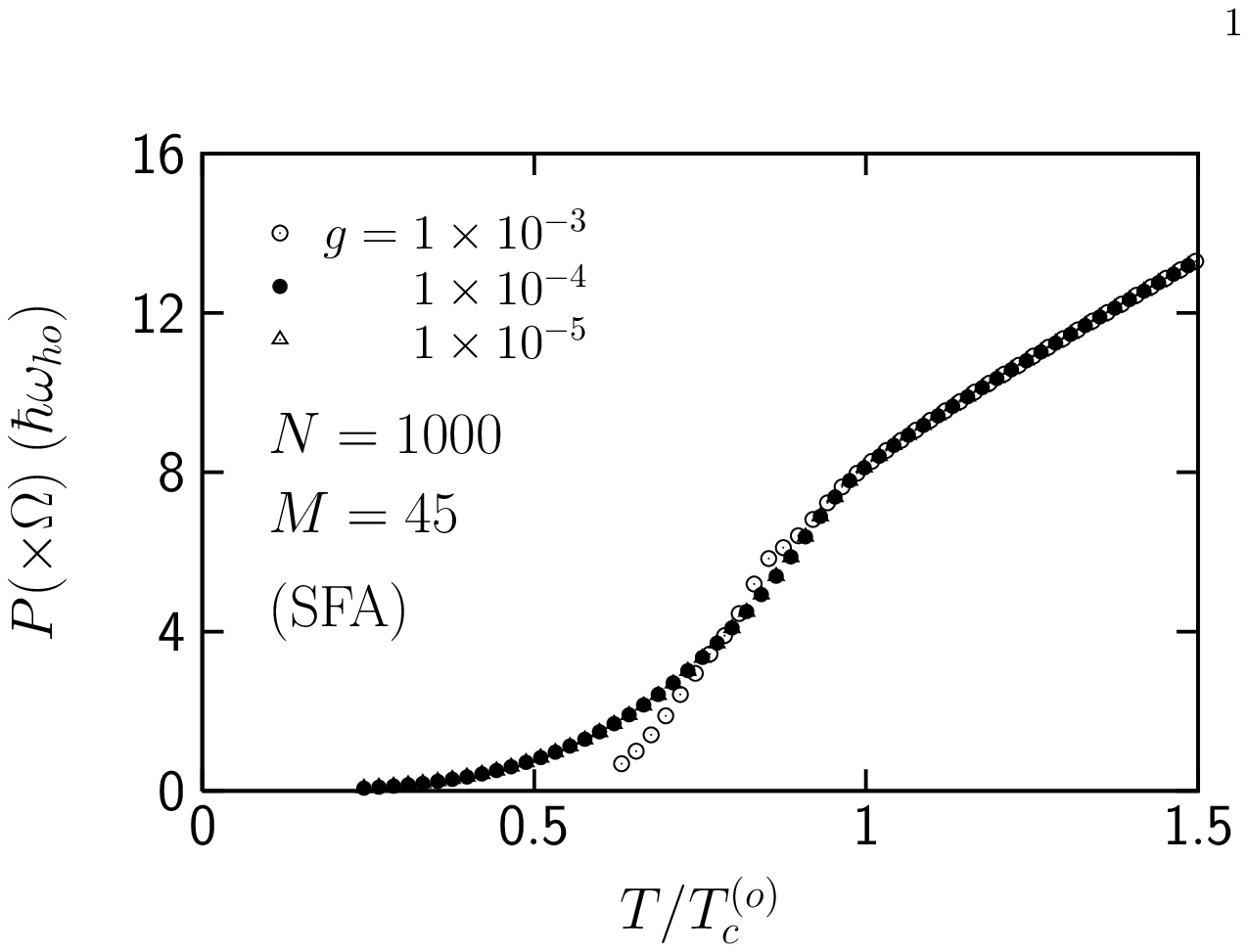}
\caption{The SFA pressure ($P\times \Omega$) as a function of $T$ for $N=1000$ trapped HS bosons,
$M=45$ states, and different $g$.} \label{fig:plot.P.vs.g.N1000}
\end{center}
\end{figure}

\begin{figure}[h!]
\begin{center}
\includegraphics[width=8.5cm,bb = 165 456 538 710,clip]{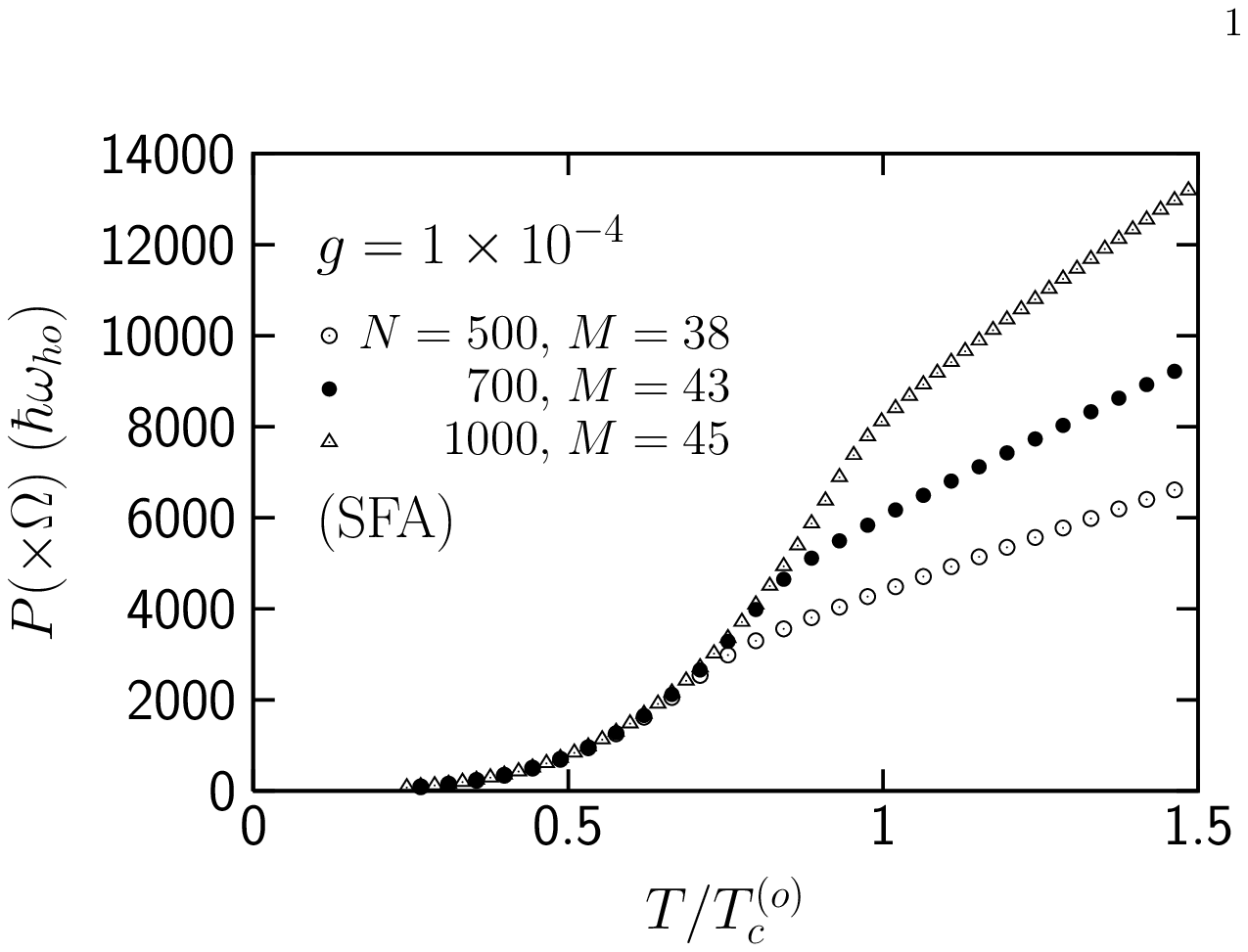}
\caption{The SFA pressure ($P\times \Omega$) as a function of $T$ for a fixed $g=1\times 10^{-4}$ 
and different $N$ and $M$ of trapped HS Bose gases.} \label{fig:plot.P.vs.N.g1e-4}
\end{center}
\end{figure}

\hs Figure\ft\ref{fig:plot.P.vs.g.N1000} displays the pressure $P (\times \Omega)$ 
as a function of $T$ for the same systems of Fig.\ft\ref{fig:plot.S.vs.g.N1000}. $P$ rises with $T$ steeply up to $T_c^{(0)}$, after which it kinks 
slightly towards a lower slope. Figure\ft\ref{fig:plot.P.vs.N.g1e-4} is the same as 
Fig.\ft\ref{fig:plot.P.vs.g.N1000}, but for fixed $g$ and various $N$. We 
note that, at lower $T$ up to $T/T_c^{(0)}\sim 0.75$, $P$ is the same for all $N$; but as one 
approaches $T_c^{(0)}$ the pressures for different $N$ begin to deviate substantially from each other. At higher $T$, the slope of $P \Omega$ vs. $T$ in 
Fig.\ft\ref{fig:plot.P.vs.g.N1000}, evaluated between $T/T_c^{(0)}=1.0$ and 1.5, is
$\sim 2.305\times 10^{12}$ K$^{-1}$. This value is very close to the classical 
value $Nk_B$ divided by the trap energy unit $\hbar\omega_{ho}$, where 
$Nk_B/\hbar\omega_{ho}\,=\,2.084\times 10^{12}$ K$^{-1}$ for 
$\omega_{ho}=2\pi \times 10$ Hz and $N=1000$ particles. The same is true for 
Fig.\ft\ref{fig:plot.P.vs.N.g1e-4}, where the slope is different for each $N$ in 
the classical limit.

\subsection{Comparison to previous work}

\hs \"Ohberg and Stenholm \cite{Ohberg:1997} presented qualitatively a numerical 
study of an inhomogeneous Bose gas by solving the Hartree-Fock equations for 
systems with repulsive interactions and restricting the number of particles, much as we do in our present study. 
They calculated the condensate density profiles and the thermally
excited parts of the density as functions of temperature, particle number, and 
interaction strength and found that, above the critical temperature, there is no condensate. 
They further noted that HF theory does not describe the Bose gas well for all T 
and that their theory breaks down near the critical temperature, contrary to SFA. In fact, we 
have shown that SFA functions well below $T_c$. 
The HF equations were solved by using an iterative approach which involves 
a density $n(\mathbf{r})$, then solving them for this density and using the new 
eigenfunctions to calculate a new density, and so on. This procedure was then iterated until 
one found a self-consistent solution for the density and eigenfunctions. The 
convergence of this method was checked by tracing the value of the ground-state energy. 
Thus, their main achievement was to obtain the density of the gas as a function of T.
We note that, as we do, \"Ohberg and Stenholm used a limited number of 
states which they did not show explicitly. In contrast to their approach, we used a 
fixed eigenfunction and obtained the thermodynamic properties by iterating the 
energy fluctuations [Eq. (\ref{eq:varphi_F_square})]. 

\hs Ketterle and van Druten \cite{Ketterle:1996} investigated BECs in traps for 
a finite number of particles in an ideal Bose gas. They found that the signature of BECs, namely a 
discontinuity in the specific heat capacity for a small number of particles, is 
very similar to an infinite number of particles; that as one tightens the confinement, 
$T_c$ rises; and that BEC is possible in 1D and 2D. Further, the population of the first
excited state for $N\rightarrow \infty$ is negligible, even in the absence of interactions.
As mentioned is Sec. \ref{sec:occupancy-of-states}, our occupation function 
$\langle \hat{n}_1\rangle/N$ agrees with that shown by these authors in 
Fig.2 of their paper; but they did not display the occupation function for 
states higher than $m=1$. In contrast, we have displayed the occupation function for
$m=50$ in Fig.\ft\ref{fig:plot.n50.MFandSFA.N1000.g1e-4}.

\hs Giorgini \ea\ \cite{Giorgini:1996} presented results for the 
temperature-dependence of the condensate fraction and the dependence of the critical 
temperature on the interaction between the bosons in a harmonic potential. They 
found that the thermal depletion of the condensate is enhanced in the presence of 
repulsive interactions, and that $T_c$ shifts to lower values with increasing repulsive 
interactions. In addition, finite-size effects do affect the thermodynamic properties; 
however, interactions play a more profound role as they lead to significant
depletion of the condensate. Further, when one increases the number of atoms, 
the interactions become more profound. Interestingly, comparing our figures displaying the thermodynamic 
properties for different $N$ and fixed $g$ to those for different $g$ and fixed $N$, 
we arrive at the same conclusion. Finally, Giorgini \ea\ argued that the shift in the 
critical temperature is the combined effect of a finite number of particles plus the 
interactions. They found that the shift $\delta T_c^{(o)}$ arising from finite-size effects is 
always negative and vanishes at large $N$.

\hs Kao and Jiang \cite{Kao:2006} provided a theoretical treatment for the 
effect of interactions on $T_c$. They noted that a finite 
number of particles usually drags a system away from the thermodynamic limit 
and that the effect of interactions is the hardest to treat among all 
other effects. This is substantiated by the present work since, as mentioned in 
Sec. \ref{sec:failure-of-sfa}, SFA breaks down at $g \ge 1\times 10^{-3}$. In particular, they referred to the work of Giorgini \ea\ 
\cite{Giorgini:1996} and noted that these authors incorporated only thermal 
effects into their calculations of the shift of $T_c$ and ignored
the effect of the condensate component. Accordingly, Kao and Jiang considered this 
effect and underlined the importance of the interaction between the condensate atoms themselves, 
since even near the transition a small fraction of atoms resides in the condensate state at the 
trap center. Inspecting our
Fig.\ft\ref{fig:plot.n0.MFandSFA.N1000.g1e-4}, we also see that at $T_c$
there is indeed a remaining small condensate fraction. The interaction between this remaining fraction
and the thermal component is considered explicitly in our work.
Further, they treated the system of condensate plus thermal component as a two-fluid 
model and showed that the energy gap between the thermal and condensate components determines 
the shift in the transition temperature. 

\hs Kao \ea\ \cite{Kao:2003} studied ideal Bose gases trapped in a generic power-law potential, 
using the grand canonical ensemble. Corrections to the thermodynamic properties due to a finite
number of particles were evaluated, such as the transition temperature of the system
as well as the condensate fraction. It was further argued that when the system has a finite 
number of particles, it is very important to consider the occupation fraction of excited states. 

\hs Napolitano \ea\ \cite{Napolitano:1997} considered a finite number of 
noninteracting bosons trapped in a three-dimensional isotropic harmonic 
oscillator trap, and evaluated the corresponding heat capacity. For this
purpose, they used the approach of Ketterle and van Druten \cite{Ketterle:1996}. 
In order to provide a signal for a phase transition $-$ a discontinuous 
heat capacity $-$ they calculated the heat capacity numerically for a fixed 
number of particles, $C_N$, which at the time their paper was published had not 
been measured. First, they evaluated $\mu(T)$ from the normalization condition 

\begin{equation}
N\,=\,\sum_{n=0}^\infty \gamma_n \eta(E_n),
\end{equation}
where $\eta(E_n)$ is the BE occupation function, $\gamma_n\,\equiv\,(n+1)(n+2)/2$ 
is the degeneracy of the HO state $n$, and $E_n\,\equiv\,n\hbar\omega_{ho}$ is the energy eigenvalue. 
For $T\,<\,T_c^0$, the ground state $n=0$ of the system is macroscopically occupied. Once the 
chemical potential had been evaluated, the other thermodynamic properties followed easily.  
To obtain $\mu(T)$, they found the roots of the equation

\begin{equation}
S(\mu,T)\,\equiv\,\sum_{n=0}^Q\,\gamma_n\,\eta(E_n)\,-\,N,\label{eq:Napolitanos-norm}
\end{equation}

for which $S(\mu,T)\,=\,0$; a technique which bears some resemblance to ours in 
Sec. \ref{sec:chemical-potential}. In our case, we adjust the value of 
$M$ (their $Q$) to give the correct classical limit of the specific heat capacity for 
noninteracting trapped Bose gases, namely, $3Nk_B$. However, in their approach, 
they ensured the convergence of $\mu(T)$ by iterating Eq.(\ref{eq:Napolitanos-norm}),
with $S(\mu,T)$ set to 0, each time increasing the number of states $Q$, calculating the 
resulting $\mu(T)$ thereof, and comparing it to that for the previous values of 
$Q$. If the new $\mu(T)$ differed from the previous value by only a small increment, 
then this procedure was likely to have converged. Thus, they kept on increasing 
$Q$ until convergence was achieved. Then they evaluated the specific heat capacity 
for several values of $N$ and showed that there was a discontinuity in the 
thermodynamic limit; whereas the discontinuity vanished in systems with 
a finite number of particles. Our results for the specific heat capacity for finite
systems in the weakly-interacting regime also reveal the absence of a discontinuity in $C_v$.
In fact, the work of Napolitano \ea\ is one of the rare investigations of trapped Bose 
gases that use a discrete-states approach.

\hs Grossmann and Holthaus \cite{Grossmann:1995} looked into the BEC of relatively few particles 
in traps and considered particularly the effects of particle number on the condensation
temperature and specific heat capacity. They evaluated the thermodynamic properties and found that 
the emergence of the $\lambda-$point in their specific heat capacity was solely due to the 
external trapping potential.

\hs Gnanapragasam \ea\ \cite{Gnanapragasam:2006} calculated the occupation 
number of atoms in the ground state of an ideal Bose gas and an interacting 
Bose gas in a trap. From this, they investigated the effects of temperature
and interaction on the condensate properties. They found that the depletion 
of the condensate rises with the rise in interaction and temperature, and 
that repulsive interactions stabilize the above systems; whereas in the case 
of attractive interactions, the systems up to certain negative 
$g$ are stabilized by the zero-point repulsive energy arising from the 
confinement. We may argue that the signature for the stability of our
systems is the convergence of SFA calculations to stable values of the fluctuations
$\varphi_F(m,T)$ and the chemical potential $\mu(T)$. In any case, we did not have
to worry about the stability issues of our systems. This is because the repulsive HS interactions
between the bosons balance the compressional forces arising from the external trapping
potential that try to squeeze the Bose gas radially towards the center of the trap.
Further, we found similarly that as the temperature and interaction are increased, 
more and more particles are depleted out of the condensate. This result is in line with
Gnanapragasam \ea\ \cite{Gnanapragasam:2006}.

\hs Grether \ea\ \cite{Grether:2003} described the HO trapping of bosons 
and fermions in lower dimensions, the goal being to understand these structures.  
Further, they considered a $d$-dimensional noninteracting boson or fermion gas trapped by 
several mutually perpendicular harmonic oscillator potentials and showed that, in the thermodynamic limit, 
the specific heat capacity for a trapped Bose gas in 3D reaches $3Nk_B$. 
They found that BEC can only occur when $d+\delta>2$, where $d$ is the dimensionality 
of the system, and $\delta$ the number of HO potentials used that are mutually perpendicular to each other. 
In summary, the authors calculated the thermodynamic properties and densities 
for ideal Bose and Fermi gases trapped in $\delta$ mutually perpendicular 
trapping potentials. In turns out that the behavior of some thermodynamic 
properties obtained here by using SFA is quite similar to theirs (Fig.2 of their 
paper).

\section{Conclusions}\label{sec:conclusions}
\hs In summary, we have presented a formal investigation of the thermodynamic properties 
of a trapped, interacting HS Bose gas of finite size in the weakly-interacting regime, using the
static fluctuation approximation (SFA). Our basic goal was to investigate the
effect of energy fluctuations on these thermodynamic properties.

\hs We find that these fluctuations are much smaller than 
the energy itself in the weakly-interacting ($g\le 10^{-3}$), low-density regime.
When the interactions are of order $g=10^{-2}$, the
energy fluctuations rise substantially and SFA breaks down in this 
highly-interacting regime. 

\hs Further, the critical temperature $T_c^{(0)}$ drops with increasing  
interactions for a fixed number of particles $N$, and rises with increasing $N$ 
at fixed interaction strength.

\hs In addition, changing $g$ in the range from 0 to $1\times 10^{-3}$ does not have any
significant effect on the thermodynamic properties; but changing $N$ while
keeping $g$ fixed has a more profound effect.

\hs We have also found that the specific heat capacity for a fixed $g$ increases 
with increasing $N$. The specific heat in the weakly-interacting regime, with 
a correctly adjusted total number of states, reaches the classical limit $3Nk_B$ 
at sufficiently high temperatures.

\appendix{}
\subappendix{Iterative procedure} \label{sec:sfa-iterative-procedure}
\hs Here we outline briefly the iterative method used to determine the 
energy fluctuations $\varphi_F(m,T)$ using SFA equations \cite{Joudeh:2005}, 
modified for a trapped HS Bose gas system. The basic point is to solve the
nonlinear equations presented below, particularly Eq.(\ref{eq:phi(m,T)^2}), 
so as to obtain stable values for $\varphi_F(m,T)$.
The procedure is constructed from a loop over all temperatures $T$, with chosen steps 
$\Delta T$, and an inner loop, conducting iterations at each $T$ to 
determine $\varphi_F(m,T)$. Hence, for each $T$:
\begin{itemize}
\item[(1)] one first assigns initial values for $\varphi_F(m,T)$ for all 
states $m$ and the true correlations between the 
number fluctuations $\langle \Delta \hat{n}_m \Delta \hat{n}_k \rangle_c$
for $m \ne k$. Then an inner loop is started which conducts an iterative 
procedure to optimize the values of $\varphi_F(m,T)$ and 
$\langle \Delta \hat{n}_m \Delta \hat{n}_k \rangle_c$.

\item[(2)] At the beginning of each iteration, the function $f(\mu,T)$ 
[Eq.\,(\ref{eq:minimization-function})] is minimized with respect 
to the chemical potential $\mu(T)$ only.

\item[(3)] Then having, determined $\mu(T)$ from the minimization, one evaluates the mean of 
the square of the number fluctuations: 

\begin{eqnarray}
&&\langle (\Delta \hat{n}_m )^2 \rangle\,=\,\eta_0(m,T) [1+\eta_0(m,T)]\,+
\,2\eta_1(m,T)\frac{1}{2} g\,\sum_{k=1}^M c(m,k) \langle \Delta \hat{n}_k 
\Delta \hat{n}_m \rangle_c, \nonumber\\
\end{eqnarray}
where $\eta_1(m,T)$ is given by (\ref{eq:eta0}) below, and $c(m,k)$ 
is the interaction matrix, Eq.\,(\ref{eq:interaction-matrix-Cnm}).

\item[(4)] The correlations between the fluctuations are updated, using 
$\langle (\Delta \hat{n}_m )^2 \rangle$ in

\begin{equation}
\langle \Delta \hat{n}_m \Delta \hat{n}_k \rangle \,=\, 
\langle (\Delta \hat{n}_m )^2 \rangle \delta_{m\,k}\,+\,
\langle \Delta \hat{n}_m \Delta \hat{n}_k \rangle_c.
\end{equation}

\item[(5)] $\langle \Delta \hat{n}_m \Delta \hat{n}_k \rangle$ is used to update 
the true correlations between the fluctuations, according to
\begin{eqnarray}
&&\langle \Delta \hat{n}_m \Delta \hat{n}_k \rangle_c \,=\,\eta_1(m,T)\,g
\sum_{i=1}^M c(m,i)\,\langle \Delta \hat{n}_i \Delta \hat{n}_k \rangle \nonumber\\
&&= \,\eta_1(m,T)\,g\,c(m,k)\,\langle (\Delta \hat{n}_k)^2 \rangle \,+\,\nonumber\\
&&\eta_1(m,T)\,g\,\sum_{i=1}^M c(m,i)\,\langle \Delta \hat{n}_i \Delta \hat{n}_k 
\rangle_c.
\end{eqnarray}

\item[(6)] The fluctuations of the energy are calculated using the updated 
correlations $\langle \Delta \hat{n}_m \Delta \hat{n}_k \rangle_c$:
\begin{equation}
[\varphi_F(m,T)]^2\,=\,\frac{g}{\eta_1(m,T)} \sum_{i=1}^M c(m,i)\,\langle \Delta 
\hat{n}_i \Delta \hat{n}_m \rangle_c \label{eq:phi(m,T)^2}
\end{equation}

\item[(7)] One goes back to step (2), using the updated $\varphi_F(m,T)$ and 
$\langle \Delta \hat{n}_m \Delta \hat{n}_k \rangle_c$, and steps (2)-(7) 
are repeated for a specified number of iterations. 

\item[(8)] At the end of the iterations, the procedure changes the temperature 
and is repeated.

\end{itemize}

\subappendix{SFA}\label{sec:sfa}

\hs In this appendix, we derive the thermodynamic properties using the static 
fluctuation approximation. We begin with the Heisenberg representation of 
the creation operator $\hat{b}_m^\dagger(\tau)$ given by

\begin{equation}
\hat{b}_m^\dagger(\tau)\,=\,\exp(\tau \hat{H})\,\hat{b}_m^\dagger(0)\,\exp(-\tau \hat{H}). 
\label{eq:b-in-Heisenberg-rep}
\end{equation}

This satisfies the equation of motion

\begin{equation}
\frac{d\hat{b}_m^\dagger}{d\tau}\,=\,\left[\hat{H},\hat{b}_m^\dagger(\tau)
\right]\,=\,\hat{E}_m\,\hat{b}_m^\dagger(\tau). \label{eq:eqn-of-motion}
\end{equation}
If we decompose the Hamiltonian $\hat{H}$ in (\ref{eq:HamiltonianSQ}) into a 
noninteracting and interacting part, $\hat{H}_0$ and $\hat{H}_{int}$, 
respectively, where $\hat{H}_0$ is the first, and $\hat{H}_{int}$ the second term on
the right-hand-side of Eq.(\ref{eq:HamiltonianSQ}), then $\hat{H}_0$ commutes with 
$\hat{b}_m^\dagger$ according to

\begin{equation}
\left[\hat{H}_0, \hat{b}_m^\dagger\right]\,=\,
\hbar\omega_{ho}(m+3/2)\,\hat{b}_m^\dagger;
\label{eq:commut.H0.bdagger}
\end{equation}

for the interacting part, we get
\begin{equation}
\left[\hat{H}_{int}, \hat{b}_m^\dagger\right]\,=\,
\frac{1}{2}g\sum_{n_1 n_2 n_3} c(n_1, n_2, n_3, m)\,\hat{b}_{n_1}^\dagger 
\hat{b}_{n_2}^\dagger \hat{b}_{n_3},
\label{eq:comm.Hint.bdagger}
\end{equation}

where 

\begin{equation}
c(n_1, n_2, n_3, m)\,=\,\int d\mathbf{r} \phi_{n_1}^*(\mathbf{r})\,
\phi_{n_2}^*(\mathbf{r})\,
\phi_{n_3}(\mathbf{r})\,\phi_m(\mathbf{r}).
\label{eq:many-body-interaction-integral}
\end{equation}

The energy $\hat{E}_m$ can then be evaluated from the commutation relation

\begin{equation}
\left[\hat{b}_m, \left[\hat{H}, \hat{b}_m^\dagger \right] \right]\,=\,
\hat{E}_m.
\label{eq:Em_from_commutation}
\end{equation}

which leads to

\begin{equation}
\hat{E}_m\,=\,\hbar\omega_{ho}(m+3/2)\,+\,\frac{1}{2}g\,\sum_{n_1 n_3} c(n_1, m, n_3, m)\,\hat{b}_{n_1}^\dagger \hat{b}_{n_3}. 
\label{eq:Em_n1n3}
\end{equation}

The average energy $\langle \hat{E}_m \rangle$ becomes, then,

\begin{eqnarray}
&&\langle \hat{E}_m \rangle \,=\,\nonumber\\
&&\hbar\omega_{ho}(m+3/2)\,+\,\frac{1}{2}g\,\sum_{n_1 n_3} c_(n_1, m, n_3, m) \langle \hat{b}_{n_1}^\dagger \hat{b}_{n_3}\rangle \delta_{n_1 n_3}, 
\nonumber\\ \label{eq:average-energy-incomplete}
\end{eqnarray}

which leads to Eq.(\ref{eq:avenergy_m}). 

\hs In SFA, we define the energy operator by 

\begin{equation}
\hat{E}_m\,=\,\langle \hat{E}_m \rangle\,+\,\Delta \hat{E}_m,
\end{equation}

where $\Delta \hat{E}_m$ is the fluctuation operator of the energy. 
The basic idea of SFA is to replace the square of the 
fluctuation operator with its mean value:
\begin{equation}
(\Delta \hat{E}_m)^2\,\approx\,\langle (\Delta \hat{E}_m)^2 \rangle = 
\varphi_m^2. \label{eq:sfa-approximation}
\end{equation}

\hs Going back to Eq.(\ref{eq:eqn-of-motion}), one can write the solution to this equation 
as

\begin{equation}
\hat{b}_m^\dagger(\tau)\,=\,\hat{b}_m^\dagger(0)\,\exp[\langle \hat{E}_m 
\rangle \tau]\,\exp(\Delta \hat{E}_m \tau).
\label{eq:solution-to-eqn-of-motion}
\end{equation}

With the aid of the identity \cite{Joudeh:2005},

\begin{equation}
B[a+b\Delta\hat{E}_m]\,\equiv\,\eta_0(m)\,+\,\eta_1(m)\,\Delta \hat{E}_m, 
\label{eq:B_identity}
\end{equation}

where
\begin{eqnarray}
&&\eta_0(m)\,\equiv\,\frac{1}{2} \left\{B[a+b\varphi_m]\,+\,B[a-b\varphi_m] \right\};
\nonumber\\
&&\eta_1(m)\,\equiv\,\frac{1}{2\varphi_m} \left\{B[a+b\varphi_m]\,-\,B[a-b\varphi_m] \right\}; \nonumber\\
\end{eqnarray}

we can rewrite Eq.(\ref{eq:solution-to-eqn-of-motion}) as follows:

\begin{eqnarray}
&&\hat{b}_m^\dagger(\tau)\,=\,\nonumber\\
&&\hat{b}_m^\dagger(0)\,\exp(\langle \hat{E}_m \rangle \tau )\,
\left[\cosh(\varphi_m\tau)\,+\,\frac{\Delta \hat{E}_m}{\varphi_m} 
\sinh(\varphi_m \tau) \right]. \nonumber\\
\label{eq:bdagger(tau)using.identity}
\end{eqnarray}

Using the identity 

\begin{equation}
\langle \hat{A}(\beta)\,\hat{B} \rangle\,=\,\langle \hat{B}\,\hat{A}\rangle\,=\,\frac{1}{Q}\,{\large\mathbf{Tr}}[\exp(-\beta\hat{H})\,
\hat{B}\,\hat{A}],
\label{eq:trace-identity}
\end{equation}
where $Q$ is the Gibbs partition function, and replacing $\beta$ 
with time $\tau$, we can obtain the so-called long-range equation \cite{Joudeh:2005}

\begin{equation}
\langle \hat{n}_m \hat{A} \rangle \,=\,\eta_0(m)\,\langle \hat{A} \rangle \,+
\,\eta_1(m)\,\langle \Delta \hat{E}_m \hat{A} \rangle, \label{eq:long-range-equation}
\end{equation}

where, by invoking the chemical potential via $\hat{H}\rightarrow \hat{H}-\mu$,

\begin{eqnarray}
&&\eta_0(m)\,\equiv\,\frac{1}{2}\left\{\frac{1}{\exp[\beta(\langle \hat{E}_m
\rangle\,-\,\mu\, +\varphi_m)]\,-\,1}\,+\,\frac{1}{\exp[\beta(\langle \hat{E}_m \rangle\,-\,\mu\,-\,\varphi_m)]\,-\,1}\right\}; \nonumber \\
&&\eta_1(m)\,\equiv\,\frac{1}{2\,\varphi_m}\left\{\frac{1}{\exp[\beta(\langle 
\hat{E}_m \rangle\,-\,\mu\,+\varphi_m)]\,-\,1}\,-\,\frac{1}{\exp[\beta(\langle \hat{E}_m \rangle\,-\,\mu\,-\,\varphi_m)]\,-\,1} \right\}.\nonumber\\
\label{eq:eta0}
\end{eqnarray}

The long-range equation is now applied to derive expressions for the 
fluctuations in the number of particles and the correlations between them. 
Substituting $\hat{A}\,=\,1$ into this equation, and using the fact 
that the quadratic fluctuations are symmetric, i.e., $\langle \Delta \hat{E}_m 
\rangle\,=\,0$, we obtain for the particle distribution

\begin{equation}
\langle \hat{n}_m \rangle\,=\,\eta_0(m). \label{eq:average_n_m}
\end{equation}

We also define the fluctuations in the number of particles 
$\Delta \hat{n}_m$ via

\begin{equation}
\Delta \hat{n}_m\,=\,\hat{n}_m\,-\,\langle \hat{n}_m \rangle, 
\label{eq:number-fluctuations}
\end{equation}

which then gives

\begin{equation}
\langle \Delta \hat{n}_m \hat{A} \rangle\,=\,\eta_1(m)\,
\langle \Delta \hat{E}_m \hat{A} \rangle.
\label{eq:average_delta_n_A}
\end{equation}

Substituting $\hat{A}=\Delta \hat{n}_k$ into the long-range equation, we 
arrive at the expression
\begin{equation}
\langle \Delta \hat{n}_m \Delta \hat{n}_k \rangle_c\,=\,\eta_1(m)\,\langle 
\Delta \hat{E}_m \Delta \hat{n}_k \rangle, 
\label{eq:delta_n_delta_n_true_correlations} 
\end{equation}

where the index $c$ denotes true correlations, i.e., $m\ne k$. 

\hs Now the fluctuations in the energy arise from the interactions between 
the particles. It is straightforward to show that

\begin{equation}
\Delta \hat{E}_m\,=\,\frac{1}{2}g\,\sum_p c_(p,m) \Delta \hat{n}_p. 
\label{eq:delta_E_eq_Hint}
\end{equation}

\hs The true correlations between the fluctuations can be decomposed into

\begin{equation}
\langle \Delta \hat{n}_p \Delta \hat{n}_k \rangle\,=\,
\langle (\Delta \hat{n}_k)^2 \rangle\,\delta_{p,k}+\,\langle \Delta \hat{n}_p 
\Delta \hat{n}_k \rangle_c.
\end{equation}

To find $\langle (\Delta \hat{n}_k)^2 \rangle$, we invoke 
Eq.(\ref{eq:trace-identity}) with $\hat{A}\,=\,\hat{b}_k^\dagger$ to get

\begin{equation}
\langle \hat{b}_k^\dagger (\beta) \hat{b}_k \hat{n}_k \rangle\,=\,
\langle \hat{b}_k \hat{n}_k \hat{b}_k^\dagger\rangle. \label{eq:average_bbn}
\end{equation}

Applying the commutation relation $\left[\hat{n}_k,
\hat{b}_k^\dagger\right]\,=\,\hat{b}_k$, we find that
\begin{equation}
\langle \hat{b}_k^\dagger (\beta) \hat{b}_k \hat{n}_k \rangle\,=\,
\langle 1\,+\,\hat{n}_k^2\,+\,2\hat{n}_k\rangle. \label{eq:1+nsq+2n}
\end{equation}

Since the energy operator $\hat{E}_m$ commutes with $\hat{b}_k^\dagger$ and 
$\hat{b}_k$, Eq.(\ref{eq:average_bbn}) can be written in the form

\begin{equation}
\langle \hat{b}_k^\dagger (\beta) \hat{b}_k \hat{n}_k \rangle \,=\,
\langle \exp[\beta(\hat{E}_k-\mu)] \hat{b}_k^\dagger \hat{b}_k \hat{n}_k)\rangle.
\label{eq:bbn_and_exp}
\end{equation}

Setting $\hat{b}_k^\dagger \hat{b}_k\,=\,\hat{n}_k$ on the right side of 
(\ref{eq:bbn_and_exp}), and equating it to the right side of 
(\ref{eq:1+nsq+2n}), we get

\begin{eqnarray} 
\langle \hat{n}_k^2 \rangle\,&=&\,\frac{1\,+\,2\hat{n}_k}
{\exp[\beta(\hat{E}_k-\mu)]\,-\,1} \nonumber\\
&=&\langle \hat{n_k} \rangle (1\,+\,2\langle \hat{n}_k \rangle)\,-
\,2\,\eta_1(k)\, \langle \hat{n}_k \Delta \hat{E}_k \rangle. \nonumber\\
\label{eq:nsq_in_another_way} 
\end{eqnarray}

From Eq.(\ref{eq:nsq_in_another_way}) one can determine the fluctuations in 
the number of particles:

\begin{eqnarray}
&&\langle (\Delta \hat{n}_k )^2 \rangle\,=\, \langle \hat{n}_k^2 \rangle\,-\,
\langle \hat{n}_k \rangle^2 \nonumber \\
&&\langle (\Delta \hat{n}_k)^2 \rangle\,=\,
\langle \hat{n}_k \rangle \left(1\,+\,\langle \hat{n}_k \rangle\right)\,+\,
2\eta_1(k)\,\frac{1}{2}g\,\sum_p c(k,p)\,\langle \Delta \hat{n}_p \Delta \hat{n}_k 
\rangle_c. \nonumber\\ \label{eq:final_nsq}
\end{eqnarray}

Setting $\hat{A}\,=\,\Delta \hat{E}$ in Eq.(\ref{eq:average_delta_n_A}), 
we obtain an expression for the fluctuations in the energy, $\varphi_m$:

\begin{equation}
\eta_1(m)\,\varphi_m^2\,=\,\langle \Delta \hat{n}_m \Delta \hat{E}_m \rangle 
\,=\,\sum_k c(k,m) \langle \Delta\hat{n}_m \Delta \hat{n}_k \rangle_c. \label{eq:varphi_F_square}
\end{equation}

\hs Next, the Gibbs partition function is derived which contains the fluctuations. This is given by

\begin{eqnarray}
&&Q\,=\,\hbox{Tr}\{\exp[-\beta\,(\hat{H}-\hat{N}\mu)]\}\,=\nonumber\\
&&\sum_{n_p} \exp\left[-\beta\sum_p (\hat{E}_p -\mu)\hat{n}_p \right] \nonumber\\
&&=\prod_p \sum_{n_p} \exp\left[-\beta\,(\hat{E}_p-\mu)\hat{n}_p\right]. \nonumber\\
\end{eqnarray}

where $\hat{N}=\sum_p \hat{N}_p$ Taking the logarithm for convenience, we get

\begin{eqnarray}
\ln Q\,&=&\,\ln\left\{\prod_p \sum_{n_p}\exp\left[-\beta\,
(\hat{E}_p-\mu)\hat{n}_p\right]\right\} \nonumber\\
&=&\,\sum_p \ln\left\{\sum_{n_p}\exp[-\beta(\hat{E}_p -\mu)\hat{n}_p]\right\} 
\nonumber\\
&=&\,\sum_p \ln\left\{\frac{1}{1\,-\,\exp\left[-\beta(\hat{E}_p-\mu)\right]}\right\}. 
\nonumber\\
\end{eqnarray}

Substituting $\hat{E}_m\,=\,\langle \hat{E}_m\rangle\,+\,\Delta \hat{E}_m$, and 
using the identity (\ref{eq:B_identity}), we arrive at

\begin{equation}
\ln Q\,=\,-\sum_p \left[q_0(p,T)\,+\,q_1(p,T)\,\Delta\hat{E}_p\right]. \label{eq:final-Q-expression-from-B_identity}
\end{equation}

Considering the symmetry of the two eigenvalues of the energy 
fluctuation operator, we finally get

\begin{eqnarray}
&&\ln Q\,=\,-\sum_p q_0(p,T)\nonumber\\
&&q_0(p)\,=\,\frac{1}{2}\,\ln\left(\left\{1\,-\,
\exp\left[-\beta(\langle \hat{E}_p \rangle\,-\,\mu\,+\,\varphi_p)\right] \right\} 
\right.\times \nonumber\\
&&\left.\left\{1\,-\,\exp\left[-\beta(\langle \hat{E}_p 
\rangle\,-\,\mu\,-\,\varphi_p)\right]\right\}\right). \label{eq:q_0(m,T)}
\end{eqnarray}

From $q_0(p,T)$ all other thermodynamic properties follow. 

\subappendix{$\ln Q$ in the limit $M\rightarrow \infty$}\label{sec:thermo-non-uniform}

Assuming, then, that 
$\phi_F(m,T)\rightarrow 0$, we write first

\begin{equation}
\ln Q\approx -\sum_{m=0}^M \ln \left\{1\,-\,\exp[-\beta(\hat{E}_m(T)-\mu(T))]\right\}.
\end{equation}

Following Grether \ea\ \cite{Grether:2003}, we evaluate the above sum by letting $M\rightarrow \infty$ and using
their Eq.(6):

\begin{equation}
\Omega(T, V, \mu)\,=\,\Omega_0\,-\,\frac{1}{\beta(\hbar\omega)^3} \sum_{\ell=1}^\infty \frac{\left[\displaystyle \exp(\beta(\mu(T)-3\hbar\omega/2))\right]^\ell}{\ell^4},
\end{equation}

where

\begin{equation}
\Omega_0\,=\,k_BT \ln\left\{1\,-\,\exp[-\beta(3\hbar\omega/2-\mu)]\right\}
\end{equation}

is the ground-state contribution to the grand potential. Here we have used BE statistics and harmonic 
trapping in 3D. Using their Eq.(7), we obtain

\begin{equation}
\Omega(T, V, \mu)\,\equiv\,\Omega_0\,-\,\frac{1}{\beta^4 (\hbar\omega)^3 \Gamma(4)} \displaystyle\int_0^\infty \frac{x^3}{z^{-1} \exp(x)-1}dx.
\end{equation}

This can further be written

\begin{equation}
\Omega(T, V, \mu)\,=\,\Omega_0\,-\,k_BT \left(\frac{k_BT}{\hbar\omega}\right)^3\,g_4(\alpha),
\end{equation}

with $\alpha\equiv(\mu-3\hbar\omega/2)/(k_BT)$. Apart from the fact that the treatment of Grether \ea\ \cite{Grether:2003}
uses a reference potential $\Omega_0$ and a reference chemical potential $3\hbar\omega/2$, this thermodynamic potential is the same as Eq.(1) in the paper of
Rochin \cite{Rochin:2005a} where the `harmonic' pressure is also given by $P\,=\,-\Omega/V$, with $V=\omega^{-3}$ $-$the `harmonic' volume. 

\hs In conclusion, then, Eq.(5) in our treatment with $\phi_F(m,T)\rightarrow 0$ and $M\rightarrow \infty$
gives the thermodynamic potential for a non-uniform trapped Bose gas. Hence all the thermal properties that follow
from it are also suited for non-uniform systems. But as stated before, we were only able to
use $M=100$ because of our computational limitations.

\bibliography{./sfa}

\begin{thebibliography}{10}

\bibitem{Holzmann:2004}
{M. Holzmann and J.-N. Fuchs and G. A. Baym and J.-P. Blaizot and F. Lalo\"e},
  {\em C. R. Physique} {\bf 5}, p.~21 (2004).

\bibitem{Kim:1999}
J.~G. Kim and E.~K. Lee, {\em J. Phys. B: At. Mol. Opt. Phys.} {\bf 32}, p.
  5575 (1999).

\bibitem{Li:2007}
M.~Li, H.~Fu, Z.~Zhang and J.~Chen, {\em Phys. Rev. A} {\bf 75}, p. 045602
  (2007).

\bibitem{Grossmann:1995}
S.~Grossmann and M.~Holthaus, {\em Phys. Lett. A} {\bf 208}, p. 188 (1995).

\bibitem{Ohberg:1997}
{P. \"Ohberg and S. Stenholm}, {\em J. Phys. B: At. Mol. Opt. Phys.} {\bf 30},
  p. 2749 (1997).

\bibitem{Grether:2003}
{M. Grether, M. Fortes, M. de Llano, J. L. del R$\acute{i}$o, F. J. Sevilla, M.
  A. Sol$\acute{i}$s, and A. A. Valladares}, {\em Eur. Phys. J. D} {\bf 23}, p.
  117 (2003).

\bibitem{Anderson:1995}
M.~H. Anderson, J.~R. Ensher, M.~R. Matthews, C.~E. Wieman and E.~A. Cornell,
  {\em Science} {\bf 269}, p. 198 (1995).

\bibitem{Liu:2000}
S.~J. Liu, G.~X. Huang, L.~Ma, S.~Zhu and H.~W. Xiong, {\em J. Phys. B: At.
  Mol. Opt. Phys.} {\bf 33}, p. 3911 (2000).

\bibitem{Ketterle:1996}
W.~Ketterle and N.~J. van Druten, {\em Phys. Rev. A} {\bf 54}, p. 656 (1996).

\bibitem{Kamchatnov:2004}
A.~M. Kamchatnov, {\em Sov. Phys. JETP} {\bf 98}, p. 908 (2004).

\bibitem{Giorgini:1996}
S.~Giorgini, L.~P. Pitaevskii and S.~Stringari, {\em Phys. Rev. A} {\bf 54}, p.
  R4633 (1996).

\bibitem{Kao:2006}
Y.-M. Kao and T.~F. Jiang, {\em Phys. Rev. A} {\bf 73}, p. 043604 (2006).

\bibitem{Kao:2003}
Y.-M. Kao, D.~H. Lin, P.~Han and P.-G. Luan, {\em Eur. Phys. J. B} {\bf 34},
  p.~55 (2003).

\bibitem{Rochin:2005}
V.~R. Rochin and V.~S. Bagnato, {\em Braz. J. Phys.} {\bf 35}, p. 607 (2005).

\bibitem{Napolitano:1997}
R.~Napolitano, J.~D. Luca, V.~S. Bagnato and G.~C. Marques, {\em Phys. Rev. A}
  {\bf 55}, p. 3954 (1997).

\bibitem{Gnanapragasam:2006}
G.~Gnanapragasam, S.-H. Kim and M.~P. Das, {\em Mod. Phys. Lett. B} {\bf 20},
  p. 1839 (2006).

\bibitem{Giorgini:2000}
S.~Giorgini, {\em Phys. Rev. A} {\bf 61}, p. 063615 (2000).

\bibitem{Joudeh:2005}
B.~R. Joudeh, M.~K. Al-Sugheir and H.~B. Ghassib, {\em Int. J. Mod. Phys. B}
  {\bf 19}, p. 3985 (2005).

\bibitem{Sandouqa:2006}
A.~S. Sandouqa, M.~K. Al-Sugheir and H.~B. Ghassib, {\em Int. J. Theor. Phys.}
  {\bf 45}, p. 152 (2006).

\bibitem{Nigmatullin:2000}
{R. R. Nigmatullin, A. A. Khamzin, and H. B. Ghassib}, {\em Phys. Rev. E} {\bf
  61}, p. 3441 (2000).

\bibitem{Al-Sugheir:2001}
M.~K. Al-Sugheir, H.~B. Ghassib and R.~R. Nigmatulin, {\em Int. J. Theor.
  Phys.} {\bf 40}, p. 1033 (2001).

\bibitem{Al-Sugheir:2002}
M.~K. Al-Sugheir and H.~B. Ghassib, {\em Int. J. Theor. Phys.} {\bf 41}, p. 705
  (2002).

\bibitem{Ghulam:2007}
N.~M. Ghulam, H.~B. Ghassib and M.~K. Al-Sugheir, {\em Phys. Rev. C} {\bf 75},
  p. 064317 (2007).

\bibitem{Sevincli:2007}
S.~Sevincli and B.~Tanatar, {\em Phys. Lett. A} {\bf 371}, p. 389 (2007).

\bibitem{Yukalov:2006}
V.~I. Yukalov, {\em Phys. Rev. A} {\bf 72}, p. 033608 (2005).

\bibitem{Press:1999}
W.~H. Press, S.~A. Teukolsky, W.~T. Vetterling and B.~B. Flannery, {\em
  {{N}umerical {R}ecipes in {C}}}, second edn. ({C}ambridge {U}niversity
  {P}ress, 1999).

\bibitem{Arfken:1995}
G.~B. Arfken and H.~J. Weber, {\em {Mathematical Methods for Physicists}},
  fourth edn. ({Academic Press, San Diego}, {USA}, 1995).

\bibitem{Rochin:2005a}
V.~R. Rochin, {\em Phys. Rev. Lett.} {\bf 94}, p. 130601 (2005).

\bibitem{Dalfovo:1999}
F.~Dalfovo, S.~Giorgini, L.~P. Pitaevskii and S.~Stringari, {\em Rev. Mod.
  Phys.} {\bf 71}, p. 463 (1999).

\bibitem{Su:2006}
G.~Su, J.~Chen and L.~Chen, {\em Physica A} {\bf 368}, p. 459 (2006).

\bibitem{Pethick:2002}
C.~J. Pethick and H.~Smith, {\em {Bose-Einstein Condensation in Dilute Gases}},
  first edn. ({Cambridge University Press}, {Cambridge UK}, 2002).

\end{thebibliography}

\end{document}